\documentclass[aps,prc,nofootinbib]{revtex4}
\usepackage{amsmath,amssymb}
\usepackage{graphicx}

\newcommand{\be}{\begin{equation}}
\newcommand{\ee}{\end{equation}}
\newcommand{\ba}{\begin{eqnarray}}
\newcommand{\ea}{\end{eqnarray}}
\newcommand{\ban}{\begin{eqnarray*}}
\newcommand{\ean}{\end{eqnarray*}}
\newcommand \nn {\nonumber}

\begin{document}

\title{Production of $^4{\rm Li}$ and $p\!-\!^3{\rm He}$ correlation function \\ in relativistic heavy-ion collisions}

\author{Sylwia Bazak$^1$\footnote{e-mail: sylwia.bazak@gmail.com} and Stanis\l aw Mr\' owczy\' nski$^{1,2}$\footnote{e-mail: stanislaw.mrowczynski@ncbj.gov.pl}}

\affiliation{$^1$Institute of Physics, Jan Kochanowski University, ul. Uniwersytecka 7, PL-25-406 Kielce, Poland 
\\
$^2$National Centre for Nuclear Research, ul. Pasteura 7,  PL-02-093 Warsaw, Poland}

\date{June 5, 2020}

\begin{abstract}

The thermal and coalescence models both describe well yields of light nuclei produced in relativistic heavy-ion collisions at LHC. We propose to measure the yield of $^4{\rm Li}$ and compare it to that of $^4{\rm He}$ to falsify one of the models. Since the masses of $^4{\rm He}$ and $^4{\rm Li}$ are almost equal, the yield of $^4{\rm Li}$ is about 5 times bigger than that of $^4{\rm He}$ in the thermal model because of different numbers of spin states of the two nuclides. Their internal structures are, however, very different: the alpha particle is well bound and compact while $^4{\rm Li}$ is weakly bound and loose. Consequently, the ratio of yields of $^4{\rm Li}$ to $^4{\rm He}$ is significantly smaller in the coalescence model and it strongly depends on the collision centrality. Since the nuclide $^4{\rm Li}$ is unstable and it decays into $^3{\rm He}$ and $p$, the yield of $^4{\rm Li}$ can be experimentally obtained through a measurement of the $p\!-\!^3{\rm He}$ correlation function. The function carries information not only about the yield of $^4{\rm Li}$ but also about the source of $^3{\rm He}$ and allows one to determine through a source-size measurement whether of $^3{\rm He}$ is directly emitted from the fireball or it is formed afterwards. We compute the correlation function taking into account the $s-$wave scattering and Coulomb repulsion together with the resonance interaction responsible for the $^4{\rm Li}$ nuclide. We discuss how to infer information about an origin of $^3{\rm He}$ from the correlation function, and finally a method to obtain the yield of $^4{\rm Li}$ is proposed. 

\end{abstract}



\maketitle

\section{Introduction}

Production of light nuclei in nucleus-nucleus collisions has been studied for decades but experimental data from Relativistic Heavy Ion Collider (RHIC)  \cite{Adler:2001uy,Agakishiev:2011ib} and Large Hadron Collider (LHC)  \cite{Adam:2015vda,Acharya:2017fvb,Acharya:2017bso} have revived an interest in the problem and attracted a lot of attention. In heavy-ion collisions at low energies, light nuclei occur as remnants of incoming nuclei. At high collision energies we also deal with a genuine production process -- the energy released in a collision is converted into masses of baryons and antibaryons which form nuclei and antinuclei. This is the only production mechanism in proton-proton collisions and the dominant mechanism in heavy-ion collisions when light nuclei are produced at midrapidity where fragments of the projectile and target do not show up. The numbers of the nuclei and antinuclei at midrapidity are approximately equal to each other at RHIC and are exactly equal at LHC. This clearly shows that the matter created in the collisions is (almost) baryonless - there is no net baryon charge. Together with deuterons and antideuterons, tritons and antitritons, $^3{\rm He}$ and $^3\overline{\rm He}$, $^4{\rm He}$ and $^4\overline{\rm He}$ there are also produced hypertritons and antihypertritons at RHIC and LHC \cite{Abelev:2010rv,Adam:2015yta}.

According to the coalescence model \cite{Butler:1963pp,Schwarzschild:1963zz} proposed over half a century ago production of light nuclei is a two step process: production of nucleons and formation of nuclei. A distinction of the two steps is well-founded if the energy scale of the first step is much higher than that of the second one. This is the case of nuclei produced at midrapidity in collider experiments. A characteristic energy scale of nucleon production is the double nucleon mass while the energy scale of formation of a nucleus is its binding energy. The very different energy scales of two steps correspond to the very different temporal scales. Therefore, one can indeed say that nucleons are produced at first and nuclei are formed later on due to final state interactions among nucleons which are close neighbors in the phase-space. 

We note that the coalescence model is much better justified at RHIC or LHC than at low energies where light nuclei occur as fragments of incoming nuclei. In the latter case there is hardly any separation of the energy scales of the two steps. Nevertheless the model is known to work well in a broad range of collision energies and thus it is not surprising that it properly describes production of light nuclei and antinuclei at LHC \cite{Sun:2015ulc,Sun:2017ooe,Zhu:2015voa,Zhu:2017zlb,Wang:2017smh}.

The thermodynamical model of particle production, see the review \cite{BraunMunzinger:2003zd}, is also more reliable and simpler at the highest available collision energies than at lower ones. At LHC thousands of hadrons are produced and thus it is easier to justify the statistical assumption of equipartition of energy. The model is also simpler because the matter created in the collisions is, as mentioned above, baryonless. Therefore, the baryon chemical potential vanishes and particle's yields are determined by their masses and spin degeneracy factors together with solely two thermodynamical parameters: the temperature and system's volume at the chemical freeze-out. The model describes very well not only the yields of hadron species measured at LHC but it describes at the same time the yields of light nuclei and hypernuclei \cite{Andronic:2010qu,Cleymans:2011pe,Andronic:2017pug}. The simplicity of the model makes its success very impressive but very puzzling as well. 

It is hard to assume that nuclei exist in a hot and dense fireball. The temperature, which at the chemical freeze-out is 156 MeV \cite{Andronic:2017pug}, is much bigger than the nuclear binding energy per nucleon which is a few MeV. The inter-particle spacing (typically 1--2 fm) is smaller than the radii of light nuclei of interest (2--3 fm). 

Since the interaction cross sections of light nuclei exceeds typical hadron-hadron cross sections, the light nuclei can be still  disintegrated after the chemical freeze-out. So, one asks why the yields of light nuclei are given by the temperature at chemical freeze-out not by that of thermal freeze-out which is 100--120 MeV.  It has been argued \cite{Xu:2018jff,Oliinychenko:2018ugs,Vovchenko:2019aoz} that the numbers of light nuclei remain approximately constant when the hadron fireball evolves from the chemical to thermal freeze-out because light nuclei are repeatedly formed and dissociated during this time interval. However, except the problem that the characteristic size of light nuclei is not much smaller than inter-hadron spacing in the fireball, one faces another difficulty here. The characteristic time of, say, deuteron formation, which is of the order of its inverse binding energy, is roughly 100 fm/$c$. So, it is hard to assume that a deuteron is repeatedly formed and dissociated in the time interval much shorter than the formation rate. Proponents of the thermal model resolve the difficulties related to the spacial and temporal scales of interest assuming that the final state nuclei originate from compact colorless states of quarks and gluons present in the fireball \cite{Andronic:2017pug}. 

The thermal and coalescence models, which are physically quite different, were observed long ago to give rather similar yields of light nuclei \cite{DasGupta:1981xx}. Recently the observation has been confirmed \cite{Zhu:2015voa,Mrowczynski:2016xqm} with a refined version of the coalescence model \cite{Sato:1981ez,Gyulassy:1982pe,Mrowczynski:1987,Lyuboshitz:1988,Mrowczynski:1992gc} which properly takes into account a quantum-mechanical character of the formation process of light nuclei. 

The question arises whether the final state formation of light nuclei can be quantitatively distinguished from the creation in a fireball. One thus asks whether the thermal approach to the production of light nuclei or the coalescence model can be falsified. 

It has been recently suggested \cite{Mrowczynski:2016xqm} and worked out in \cite{Bazak:2018hgl} to compare the yield of $^4{\rm He}$ already measured at RHIC \cite{Agakishiev:2011ib} and LHC \cite{Adam:2015vda} to the yield of exotic nuclide $^4{\rm Li}$ which was discovered in Brekeley in 1965 \cite{Cerny-1965}. Since the mass of $^4{\rm He}$ is smaller than that of $^4{\rm Li}$ by only 20 MeV, the yield of $^4{\rm Li}$, which has spin 2, is according to the thermal model about five times bigger than that of $^4{\rm He}$ because of five spin states of $^4{\rm Li}$ and only one of $^4{\rm He}$. Taking into account the mass difference of $^4{\rm Li}$ and $^4{\rm He}$, the factor is reduced from 5 to 4.3 at the temperature of chemical freeze-out of 156 MeV. 

The alpha particle is well bound and compact while the nuclide $^4{\rm Li}$ is weakly bound and loose. Consequently,  the coalescence model predicts the ratio of the yields of $^4{\rm Li}$ to $^4{\rm He}$ which is significantly smaller than 5 but more importantly the ratio of the yields strongly depends, in contrast to the thermal model, on the fireball size that is on the collision centrality, see Sec.~\ref{sec-formation}. This is the distinctive feature of the coalescence model. We note that a search of $^4{\rm Li}$ production at RHIC has been recently advocated in \cite{Xi:2019vev}.

The nuclide $^4{\rm Li}$ is unstable and it decays into $p+{^3{\rm He}}$ with the width of 6 MeV \cite{NNDC}, see also \cite{Tilley:1992zz}. Since the lifetime of $^4{\rm Li}$ is about $30~{\rm fm}/c$ its yield can be experimentally obtained through the $p\!-\!^3{\rm He}$ correlation function which was measured in $^{40}{\rm Ar}-$induced reaction on $^{197}{\rm Au}$ at laboratory energy 60 MeV per nucleon \cite{Pochodzalla:1987zz}. The correlation function was also measured at relativistic collision energies at AGS and the yield of $^4{\rm Li}$ was estimated  \cite{Armstrong:2001mr}. 

The aim of this study is to compute the $p\!-\!^3{\rm He}$ correlation function which is needed to obtain the yield of $^4{\rm Li}$. However, it is not straightforward to infer the yield from a measured correlation function because the resonance, which corresponds to $^4{\rm Li}$, is not the only interaction channel  of $^3{\rm He}$ and $p$ at low relative momenta. There is also the $s-$wave scattering and Coulomb repulsion. Therefore, the resonance peak of $^4{\rm Li}$ is strongly deformed. We propose a special procedure to obtain the yield of $^4{\rm Li}$. 

When $p\!-\!D$ or $p\!-\!^3{\rm He}$ correlation functions are computed the problem is how to choose a source function of light nuclei. It has been observed in our recent paper \cite{Mrowczynski:2019yrr} that the proton-deuteron correlation function gets a different form in dependence whether deuterons are directly emitted from the fireball as all other hadrons or deuterons are formed later on due to final state interactions. Specifically, there are different source functions of deuterons, which enter the formulas of the $p\!-\!D$ correlation function, in the two cases. Consequently, the source radii inferred from the correlation functions differ by the factor $\sqrt{4/3}$.  There is an analogous situation when the $p\!-\!^3{\rm He}$ correlation function is considered but the computation presented in Sec.~\ref{sec-gen-corr-fun} is more complicated because we deal with a four-body problem. If one assumes that $^3{\rm He}$ is emitted directly from the fireball the source radius inferred from the correlation function is smaller by the factor $\sqrt{3/2}$ than that corresponding to the scenario where nucleons emitted from the fireball form the nuclide $^3{\rm He}$ due to final state interactions. When the $p\!-\!D$ or $p\!-\!^3{\rm He}$ correlation function is computed, one should adopt one of the two scenarios of light nuclei production and choose an appropriate source function. However, there is also a much more interesting consequence of our finding: knowing the nucleon source radius from the proton-proton correlation function, which can be precisely measured, see \cite{Acharya:2018gyz,Adam:2015vja}, we can quantitatively distinguish the emission of a nucleus from the fireball from the formation of the nucleus afterwards. 

The $p\!-\!^3{\rm He}$ correlation function has been recently computed \cite{Xi:2019vev}. However, the resonance interaction has not been taken into account in the correlation function but the $p\!-\!^3{\rm He}$ pairs coming from the two-body decays of $^4{\rm Li}$ have been generated by means of a Monte Carlo method and added to the correlation function which includes the $s-$wave scattering and Coulomb repulsion. Therefore, there is no interference of the incoming and outgoing waves modified by the resonance interaction and the  resonance contribution to the correlation function does not depend on the source radius. The computation \cite{Xi:2019vev} is not only oversimplified but the resonance peak seems to be far too narrow. Consequently the shape of the correlation function is very different from that of the measured function \cite{Pochodzalla:1987zz}. The $p\!-\!^3{\rm He}$ correlation functions presented in \cite{Xi:2019vev} are further compared with ours at the end of Sec~\ref{sec-corr-fun-comp}. 

Although the main objective of this paper is the $p\!-\!^3{\rm He}$ correlation function we repeat to some extend our considerations from \cite{Bazak:2018hgl} where the formation of $^4{\rm He}$ and $^4{\rm Li}$ was studied. The repetition is not only for the completeness of the present paper. We have refined some arguments and added a figure. It is also notable that some formulas introduced in the context of formation of $^4{\rm He}$ and $^4{\rm Li}$ are used in our subsequent considerations. 

The paper is organized as follows. The formation of light nuclei, in particular $^4{\rm He}$ and $^4{\rm Li}$, is discussed in Sec.~\ref{sec-formation}. In Sec.~\ref{sec-gen-corr-fun} we derive a general formula of the $p\!-\!^3{\rm He}$ correlation function which is considered first as a two and then as a four body problem. In the first case one assumes that the nuclides $^3{\rm He}$ are emitted directly from the fireball and in the second one that the nuclides are formed afterwards. In Sec.~\ref{sec-corr-fun-comp} the $p\!-\!^3{\rm He}$ correlation function is computed taking into account the resonance interaction, $s-$wave scattering and Coulomb repulsion. We discuss how to infer the information about an origin of $^3{\rm He}$ from the correlation function, and finally in Sec.~\ref{sec-yield} we propose a method to obtain the yield of  $^4{\rm Li}$. The paper is closed with the summary of our results and conclusions.

\section{Formation rate of light nuclei}
\label{sec-formation}

According to the coalescence model \cite{Butler:1963pp,Schwarzschild:1963zz}, the momentum distribution of a final state nucleus of $A$ nucleons is expressed through the nucleon momentum distribution as
\be
\label{A-mom-dis}
\frac{dP_A}{d^3p_A} = {\cal A}_A \bigg(\frac{dP_N}{d^3p} \bigg)^A,
\ee
where ${\bf p}_A = A{\bf p}$ and ${\bf p}$ is assumed to be much bigger than the characteristic internal momentum of a nucleon in the nucleus of interest, the quantity ${\cal A}_A$, which we call the {\it coalescence} or {\it formation rate}, is related to the probability that $A$ nucleons fuse into the nucleus. It is of the dimension $p^{3(A-1)}$ with $p$ being a momentum. As first derived in \cite{Sato:1981ez} and later on repeatedly discussed \cite{Gyulassy:1982pe,Mrowczynski:1987,Lyuboshitz:1988,Mrowczynski:1992gc}, the formation rate can be expressed as
\ba 
\label{A-form-rate}
{\cal A}_A  &=& g_S g_I (2\pi)^{3(A-1)} V
\int d^3r_1 \, d^3r_2 \dots d^3r_A 
D ({\bf r}_1)\, D ({\bf r}_2) \dots D ({\bf r}_A)\, |\Psi({\bf r}_1,{\bf r}_2, \dots {\bf r}_A ) |^2 ,
\ea
where $g_S$ and $g_I$ are the spin and isospin factors to be discussed later on; the multiplier $(2\pi)^{3(A-1)}$ results from our choice of natural units where $\hbar =1$; $V$ is the normalization volume which disappears from the final formula; the source function $D ({\bf r})$ is the normalized to unity position distribution of a single nucleon at the kinetic freeze-out and $\Psi({\bf r}_1,{\bf r}_2, \dots {\bf r}_A )$ is the wave function of the nucleus of interest. 

The formula (\ref{A-mom-dis}) does not assume, as one might think, that the nucleons are emitted simultaneously. The vectors ${\bf r}_i$ with $i=1,2, \dots A$ denote the nucleon positions at the moment when the last nucleon is emitted from the fireball. For this reason, the function $D ({\bf r}_i)$ actually gives the space-time distribution and it is usually assumed to be Gaussian. We choose the isotropic form
\be
\label{Gauss-source}
D ({\bf r}_i) = (2 \pi R_s^2)^{-3/2} \, e^{-\frac{{\bf r}_i^2}{2R^2_s}},
\ee
where $\sqrt{3} R_s$ is the root mean square (RMS) radius of the nucleon source. If the time duration $\tau$ of the emission is explicitly taken into account it enlarges the effective radius of the source from $R_s$ to $\sqrt{R_s^2 + v^2\tau^2}$ where $v$ is the velocity of the particle pair relative to the source. 

The Gaussian parameterization of the source function (\ref{Gauss-source}) is obviously much simpler than the more realistic blast-wave parameterization used in {\it e.g.} \cite{Zhu:2017zlb}. However, our aim is to perform analytic calculations of both the formation rates of $^4{\rm He}$ and $^4{\rm Li}$ and of the $p\!-\!^3{\rm He}$ correlation function. With the blast-wave parameterization such calculations would be very difficult if possible at all. Nevertheless, it should be stressed that the Gaussian parametrization is not only convenient for analytical calculations but there is an empirical argument in favor of this choice. The imaging technique \cite{Brown:1997ku} allows one to infer the source function from a two-particle correlation function provided the inter-particle interaction is known. The technique applied to experimental data from relativistic heavy-ion collisions showed that the source functions are mostly Gaussian and non-Gaussian contributions are rather small, see \cite{Alt:2008aa}. Let us also note that the isotropic Gaussian source function (\ref{Gauss-source}) is frequently used, in particular when one studies correlations of particles like $\Lambda$ or $\bar{\Lambda}$ which are not so copiously produced as pions or kaons, see {\it e.g.} \cite{Acharya:2019ldv}. Then, the correlation functions are not measured precisely enough to disentangle temporal and different spatial sizes of the source which are encoded in a source function more realistic than (\ref{Gauss-source}). 

The source function depends in general on particle's mass and its momentum. Actually, the source radius scales with the particle's transverse mass $m_\perp \equiv \sqrt{m^2 + p_\perp^2}$.  For the case of one-dimensional analysis relevant for our study, the effect is well seen in Fig.~8 of \cite{Adam:2015vja} where experimental data on Pb-Pb collisions at LHC are shown. The dependence of the source radius on $m_\perp$ is evident when we deal with pions and $m_\perp \lesssim 0.9~{\rm GeV}$ but it is much weaker for protons when $m_\perp \gtrsim 1.0~{\rm GeV}$. If one considers a correlation function of particles from a sufficiently small transverse momentum interval, one can use the source function (\ref{Gauss-source}) which implicitly depends on $m_\perp$ trough the source radius $R_s$.

The spin and isospin factors $g_S$ and $g_I$, which enter the formation rate (\ref{A-form-rate}), give a probability that spin and isospin quantum numbers of $A$ nucleons match the quantum numbers of the nucleus of $A$ nucleons under consideration. To compute the factors we assume that the nucleons are unpolarized with respect to spin and isospin. However, the spin and isospin are treated somewhat differently. If the spin of the nucleus is $S$, the factor $g_S$ equals $2S +1$ divided by the number of states of $A$ nucleons with the total spin $S$. In case of isospin, one must remember that a total isospin $I$ and its third component $I_3$ are assigned to a given nucleus. Therefore,  the factor $g_I$ equals the inverse number of states of $A$ nucleons with $I$ and $I_3$ of the nucleus of interest. 

To formulate a relativistically covariant coalescence model one usually uses the Lorentz invariant nucleon momentum distributions in the relation analogous to (\ref{A-mom-dis}) and modifies the coalescence rate formula (\ref{A-form-rate}), see e.g. \cite{Sato:1981ez,Mrowczynski:1987}. Since we are interested mostly in the ratio of the coalescence rates of $^4{\rm Li}$ and $^4{\rm He}$, our final result is insensitive to these heuristic modifications which are anyway not well established, as the relativistic theory of strongly interacting bound states is not fully developed. 

In the center-of-mass frame of $A-$nucleons the formula (\ref{A-form-rate}) can be treated as nonrelativistic even so momenta of nucleons are relativistic in both the rest frame of the source and in the laboratory frame. The point is that the formation rate is non-negligible only for small relative momenta of the nucleons. Therefore, the relative motion can be treated as nonrelativistic and the corresponding wave function is a solution of the Schr\"odinger equation. The source function, which is usually defined in the source rest frame, needs to be transformed to the center-of-mass frame of the pair as discussed in great detail in \cite{Maj:2009ue}. 

Let us move to the computation of the formation rate of $^4{\rm He}$. The modulus squared of the wave function of $^4{\rm He}$ is chosen as
\be
\label{alpha-wave-fun}
|\Psi_{\rm He}({\bf r}_1,{\bf r}_2, {\bf r}_3, {\bf r}_4) |^2 = C_\alpha 
e^{- \alpha ({\bf r}_{12}^2 + {\bf r}_{13}^2 + {\bf r}_{14}^2 + {\bf r}_{23}^2 + {\bf r}_{24}^2 + {\bf r}_{34}^2)},
\ee
where $C_\alpha$ is the normalization constant,  ${\bf r}_{ij} \equiv {\bf r}_i - {\bf r}_j$ and $\alpha$ is the parameter to be related to the RMS radius of $^4{\rm He}$ which is denoted as $R_\alpha$. We further use the Jacobi variables defined as
\be 
\label{Jacobi-4}
\left\{ \begin{array}{ll}
{\bf R} \equiv \frac{1}{4}({\bf r}_1+{\bf r}_2+{\bf r}_3 + {\bf r}_4) ,
\\[2mm]
{\bf x} \equiv {\bf r}_2-{\bf r}_1,
\\[2mm]
{\bf y} \equiv {\bf r}_3-\frac{1}{2}({\bf r}_1+{\bf r}_2) ,
\\[2mm]
{\bf z} \equiv {\bf r}_4 -\frac{1}{3}({\bf r}_1+{\bf r}_2+{\bf r}_3) ,
\end{array} \right.
~~~~~~~~~~~~~~~~~~
\left\{ \begin{array}{ll}
{\bf r}_1 = {\bf R} -\frac{1}{2} {\bf x} - \frac{1}{3} {\bf y} - \frac{1}{4} {\bf z} ,
\\[2mm]
{\bf r}_2 = {\bf R} + \frac{1}{2}{\bf x} - \frac{1}{3} {\bf y} - \frac{1}{4} {\bf z} ,
\\[2mm]
{\bf r}_3 = {\bf R} + \frac{2}{3} {\bf y}- \frac{1}{4} {\bf z} ,
\\[2mm]
{\bf r}_4 = {\bf R} + \frac{3}{4} {\bf z} ,
\end{array} \right.
\ee
which have the nice property that the sum of squares of particles' positions and the sum of squares of differences of the positions are expressed with no mixed terms of the Jacobi variables that is
\ba
\label{Jacobi-property-1}
{\bf r}_1^2 + {\bf r}_2^2 + {\bf r}_3^2 + {\bf r}_4^2 
= 4 {\bf R}^2 + \frac{1}{2} {\bf x}^2 + \frac{2}{3} {\bf y}^2 + \frac{3}{4} {\bf z}^2 ,
\\ \label{Jacobi-property-2}
{\bf r}_{12}^2 + {\bf r}_{13}^2 + {\bf r}_{14}^2 + {\bf r}_{23}^2 + {\bf r}_{24}^2 + {\bf r}_{34}^2 = 2 {\bf x}^2 + \frac{8}{3} {\bf y}^2 + 3 {\bf z}^2 .
\ea
Using the relation (\ref{Jacobi-property-2}), one easily finds that 
\be
C_\alpha = \frac{2^6}{V} \Big(\frac{\alpha}{\pi}\Big)^{9/2}, 
~~~~~~~~~~ 
\alpha = \frac{3^2}{2^5 R_\alpha^2}  .
\ee

Substituting the formulas (\ref{Gauss-source}) and (\ref{alpha-wave-fun}) into Eq.~(\ref{A-form-rate}), one finds the coalescence rate of  $^4{\rm He}$ as
\be
\label{alpha-rate}
{\cal A}_4^{\rm He} =  \frac{\pi^{9/2}}{2^{9/2}}
\frac{1}{\big(R_s^2 +\frac{4}{9} R_\alpha^2\big)^{9/2}} ,
\ee 
where the spin and isospin factors have been included. Since $^4{\rm He}$ is the state of zero spin and zero isospin, the factors are 
\be
g_S = g_I = \frac{1}{2^3}, 
\ee
because there are $2^4$ spin and $2^4$ isospin states of four nucleons and there are two zero spin and two zero isospin states. The coalescence rate of $^4{\rm He}$ was computed long ago in \cite{Sato:1981ez}.

The stable isotope $^6{\rm Li}$ is a mixture of two cluster configurations $ ^4{\rm He}\!-\! ^2{\rm H}$ and $ ^3{\rm He}\!-\! ^3{\rm H}$ \cite{Bergstrom:1979gpv}. Since $^4{\rm Li}$ decays into $^3{\rm He}+p$, we assume that it has the cluster structure $ ^3{\rm He}\!-\!p$ and following \cite{Bergstrom:1979gpv} we parametrize the modulus squared of the wave function of $^4{\rm Li}$ as
\ba
\label{Li-wave-fun}
|\Psi_{\rm Li}({\bf r}_1,{\bf r}_2, {\bf r}_3, {\bf r}_4) |^2 = 
C_{\rm Li} \, e^{-\beta ({\bf r}_{12}^2 + {\bf r}_{13}^2 + {\bf r}_{23}^2 )} 
{\bf z}^4 e^{-\gamma {\bf z}^2} \, |Y_{lm}(\Omega_{\bf z})|^2,
\ea
where the nucleons number 1, 2 and 3 form the $^3{\rm He}$ cluster while the nucleon number 4 is the proton; ${\bf z}$ is the Jacobi  variable (\ref{Jacobi-4}); $Y_{lm}(\Omega_{\bf z})$ is the spherical harmonics related to the rotation of the vector ${\bf z}$ with quantum numbers $l,m$. The summation over $m$ is included in the spin factor $g_S$. The nuclide $^4{\rm Li}$ is treated here as a stable one and consequently the normalization constant $C_{\rm Li}$ is a time-independent real number.  

Using the Jacobi variables, one analytically computes the constant $C_{\rm Li}$ and expresses the parameter $\beta$, which enters the formula (\ref{Li-wave-fun}), through the RMS radius $R_c$ of the cluster $^3{\rm He}$ as
\be
C_{\rm Li} = \frac{2^4 3^{1/2} \beta^3 \gamma^{7/2}}{5 \pi^{7/2} V}, ~~~~~~~~~ 
\beta = \frac{1}{3 R_c^2}  .
\ee
The parameter $\gamma$  is expressed through the RMS radius $R_{\rm Li}$ of $^4{\rm Li}$ and the cluster radius $R_c$ in the following way
\be
\gamma = \frac{21}{2^3(4R_{\rm Li}^2 - 3R_c^2)} .
\ee

Let us now discuss the spin and isopsin factors which enter the coalescence rate of  $^4{\rm Li}$. The nuclide has the isospin $I =1,~I_z = 1$ and thus the isospin factor is
\be
\label{gI-Li}
g_I = \frac{3}{2^4},
\ee
because there are three isospin states $I =1,~I_z = 1$ of four nucleons. 

The spin of the ground state of $^4{\rm Li}$ is 2 which can be arranged with the orbital angular momentum $l=1$ and $l=2$. We assume here that the cluster $^3{\rm He}$ has spin $1/2$ as the free nuclide  $^3{\rm He}$. (If the spin $3/2$ of $^3{\rm He}$ were allowed, the orbital number $l=0$ would be also possible.) When the spins of $^3{\rm He}$ and $p$ are parallel or antiparallel, the orbital number is $l=1$ or $l=2$, respectively. However, the ground state of $^4{\rm Li}$ is of negative parity which suggests that $l=1$. Indeed, the parity of a two-particle system is $P= \eta_1 \eta_2 (-1)^l$ where $\eta_1,~\eta_2$ are internal parities of the two particles. Since the parities of  ${^1{\rm H}}$ and ${^3{\rm He}}$ are both positive, the orbital momentum $l$ must be odd. Therefore, we assume further on that  $l=1$.

When $l=1$, the total spin of $^3{\rm He}$ and $p$ has to be one and there are $3^2$ such spin states of four nucleons. Consequently, there are $3^2$  angular momentum states with 5 states corresponding to spin 2 of  $^4{\rm Li}$ and thus
\be
\label{gS-l1}
g_S = \frac{3^2}{2^4} \frac{5}{3^2} = \frac{5}{2^4}.
\ee

Substituting the formulas (\ref{Gauss-source}) and (\ref{Li-wave-fun}) into Eq.~(\ref{A-form-rate}), one finds the coalescence rate of  $^4{\rm Li}$ as
\be
\label{Li-rate}
{\cal A}_4^{\rm Li} 
= \frac{15 \pi^{9/2}}{2^{13/2}}  
 \frac{R_s^4}{\big(R_s^2 +\frac{1}{2} R_c^2\big)^3 
\big(R_s^2 +\frac{4}{7} R_{\rm Li}^2 - \frac{3}{7} R_c^2\big)^{7/2}} .
\ee 
Since the source function (\ref{Gauss-source}) is spherically symmetric, the coalescence rate (\ref{Li-rate}) depends on the orbital numbers $l$ only through the spin factor $g_S$. 

\begin{figure}[t]
\begin{minipage}{87mm}
\centering
\includegraphics[scale=0.285]{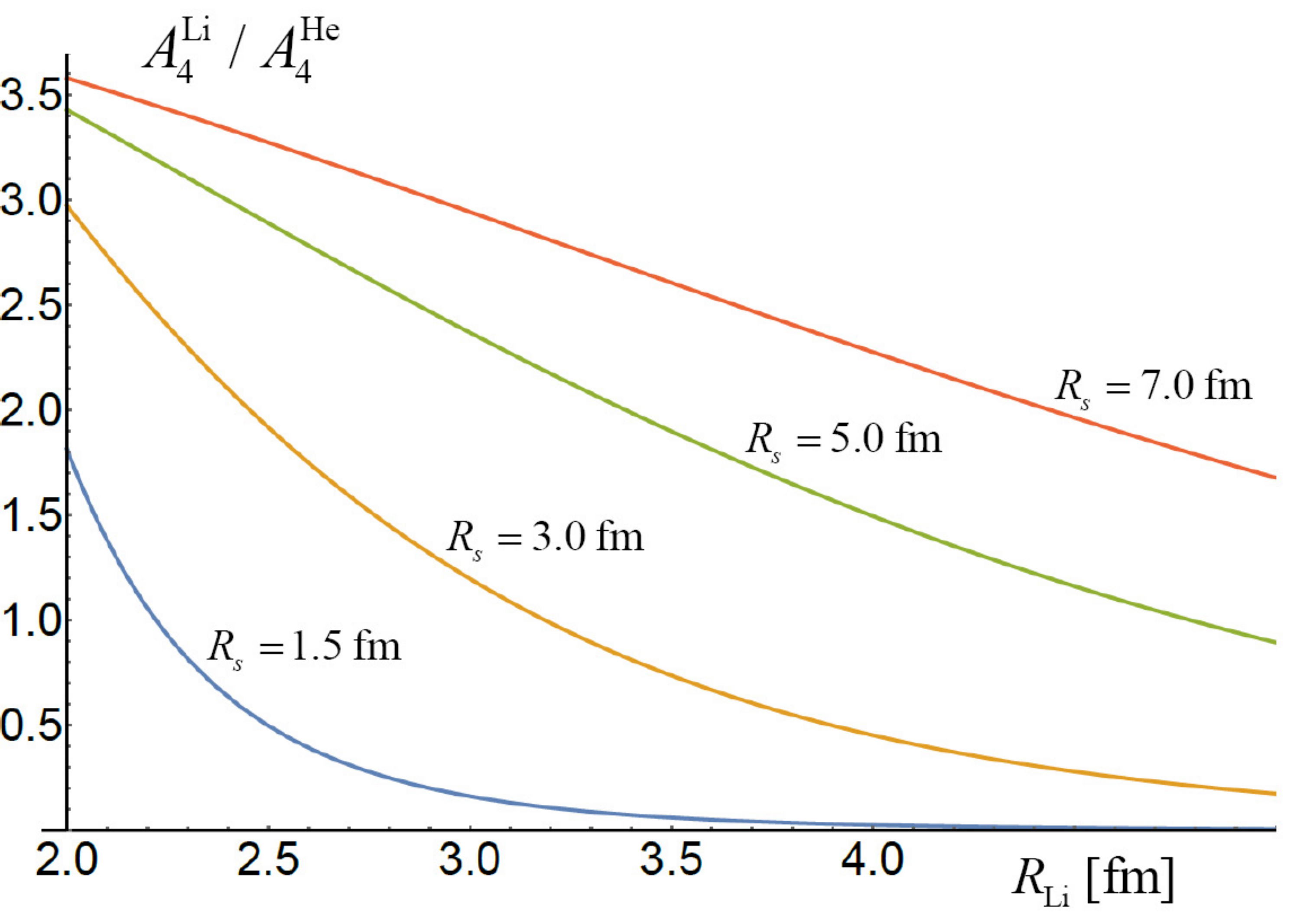}
\vspace{-5mm}
\caption{The ratio of formation rates of $^4{\rm Li}$ to $^4{\rm He}$ as a function of $R_{\rm Li}$ for four values of $R_s = 1.5,~3.0,~5.0$ and 7.0 fm.}
\label{Fig-ratio-rates-1}
\end{minipage}
\hspace{2mm}
\begin{minipage}{87mm}
\centering
\vspace{1mm}
\includegraphics[scale=0.28]{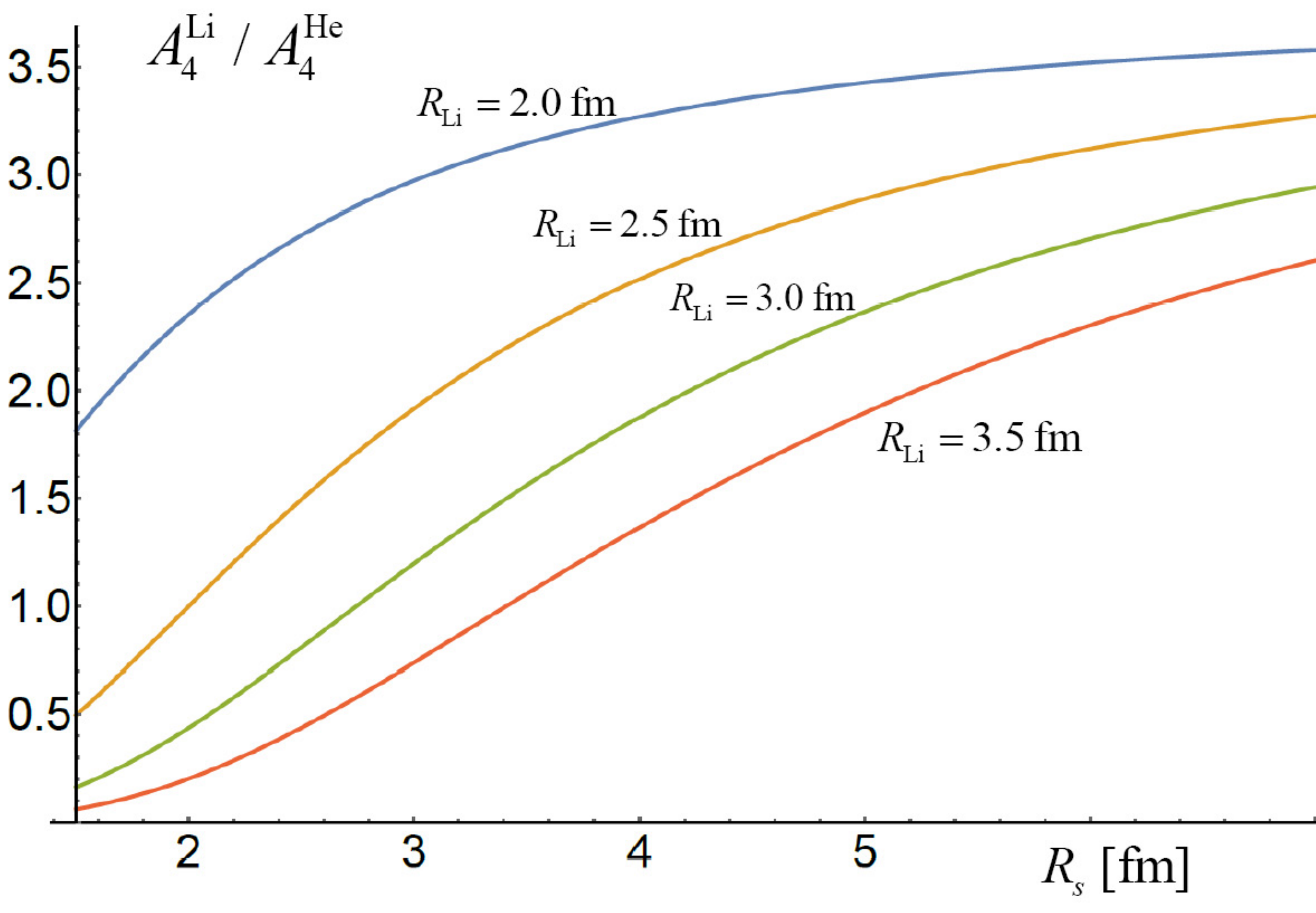}
\vspace{-6mm}
\caption{The ratio of formation rates of $^4{\rm Li}$ to $^4{\rm He}$ as a function of $R_s$ for four values of $R_{\rm Li} = 2.0,~2.5,~3.0$ and 3.5 fm.}
\label{Fig-ratio-rates-2}
\end{minipage}
\hspace{2mm}
\end{figure}

We note that even when $R_\alpha = R_{\rm Li}$ and the spin-isospin factors are ignored, the coalescence rates of $^4{\rm He}$ and   $^4{\rm Li}$ still differ from each other because the internal structure of $^4{\rm He}$ differs from that of  $^4{\rm Li}$. The rates become equal when $R_s \gg R_\alpha $ and $R_s \gg R_{\rm Li}$ as then the structure of nuclei does not matter any more. One checks that our  formulas indeed confirm the expectation. 

The ratio of yields of $^4{\rm Li}$ and $^4{\rm He}$ is given by the ratio of the formation rates ${\cal A}_4^{\rm Li}$ and ${\cal A}_4^{\rm He}$. The latter ratio depends on four parameters: $R_s$, $R_\alpha$, $R_{\rm Li}$ and $R_c$. The fireball radius at the kinetic freeze-out $R_s$ is usually determined by the femtoscopic correlations of pions which are abundantly produced. Specifically, the experimentally measured radii $R_{\rm out},\, R_{\rm side},\, R_{\rm long}$ can be used to get the kinetic freeze-out radius as $R_s = (R_{\rm out} R_{\rm side} R_{\rm long})^{1/3}$. Then, the source radius $R_s$ varies from low-multiplicity proton-proton to central Pb-Pb collisions at LHC between, say, 1 and 7 fm \cite{Aamodt:2011kd,Adam:2015vna}. For our purpose it is more appropriate to use the source radii $R_s$ obtained from the proton-proton correlation functions which have been also precisely measured at LHC \cite{Acharya:2018gyz,Adam:2015vja}.

 The RMS radius of $^4{\rm He}$ is $R_\alpha = 1.68$ fm \cite{Angeli:2013epw} and the RMS radius of the cluster $^3{\rm He}$ is identified with the radius of a free nucleus $^3{\rm He}$ and thus $R_c = 1.97$ fm \cite{Angeli:2013epw}. The radius $R_{\rm Li}$ is unknown but obviously it must be bigger than $R_c$. Taking into account a finite size of a proton it is fair to expect that $R_{\rm Li}$ is about 2.5--3.5 fm. The ratio of the formation rates ${\cal A}_4^{\rm Li}$ and ${\cal A}_4^{\rm He}$ is shown in Fig.~\ref{Fig-ratio-rates-1}  as a function of $R_{\rm Li}$ for four values of $R_s = 1.5,~3.0,~5.0$ and 7.0 fm and in Fig.~\ref{Fig-ratio-rates-2} as a function of $R_s$ for four values of $R_{\rm Li} = 2.0,~2.5,~3.0$ and 3.5 fm.

As already mentioned, the ratio of yields of $^4{\rm Li}$ and $^4{\rm He}$ equals 5 according to the thermal model if one ignores the mass difference of the nuclides. The magnitude of the ratio is reduced to 4.3 when the mass difference is taken into account and the temperature of chemical freeze-out equals 156 MeV. Figs.~\ref{Fig-ratio-rates-1} and \ref{Fig-ratio-rates-2} show that the ratio is significantly smaller in the coalescence model. For $R_{\rm Li}= 3$ fm and the most central collisions of the heaviest nuclei, which corresponds to  $R_s \approx 7$ fm, the ratio ${\cal A}_4^{\rm Li}/{\cal A}_4^{\rm He}$ equals about 3 but it drops below 2 for the centrality of 40-60\% where $R_s \approx 4$ fm. The ratio is even smaller for truly peripheral nucleus-nucleus or $p$-$p$ collisions. The strong dependence of the ratio of the yields of $^4{\rm Li}$ to $^4{\rm He}$ on the collision centrality is a characteristic feature of the coalescence mechanism. Therefore, it should be possible to quantitatively distinguish the coalescence mechanism of light nuclei production from the creation in a fireball. 

The yield of $^4{\rm Li}$ can be experimentally obtained through a measurement of the $p\!-\!^3{\rm He}$ correlation function which is discussed in the remaining part of the paper.

\section{General formula of $p\!-\!^3{\rm He}$ correlation function}
\label{sec-gen-corr-fun}

A general form of the $p\!-\!^3{\rm He}$ correlation function depends on whether $^3{\rm He}$ nuclei are emitted from a source as all other hadrons or the nuclei are formed due to final state interactions among emitted nucleons. In the first case the nuclide $^3{\rm He}$ can be treated as an elementary particle and in the second one as a bound state of three nucleons. The two cases are discussed in the two subsequent sections 

\subsection{$^3{\rm He}$ treated as an elementary particle}

When the nucleus $^3{\rm He}$ is treated as an elementary particle emitted from a source, the $p\!-\!^3{\rm He}$ correlation function is defined as
\be
\label{corr-fun-def}
\frac{dP_{p{^3{\rm He}}}}{d^3p_p  d^3p_{^3{\rm He}}} = \mathcal{R}({\bf p}_p , {\bf p}_{^3{\rm He}}) \, \frac{dP_p}{d^3p_p}  \frac{dP_{^3{\rm He}}}{d^3p_{^3{\rm He}}} ,
\ee
where $\frac{dP_p}{d^3p_p}$,  $ \frac{dP_{^3{\rm He}}}{d^3p_{^3{\rm He}}}$ and $\frac{dP_{p{^3{\rm He}}}}{d^3p_p  d^3p_{^3{\rm He}}}$ are probability densities to observe $p$, ${^3{\rm He}}$ and  $p\!-\!^3{\rm He}$  pair with momenta ${\bf p}_p$, ${\bf p}_{^3{\rm He}}$ and $({\bf p}_p, {\bf p}_{^3{\rm He}})$. If the correlation is due to final state interactions, the correlation function is known to be \cite{Koonin:1977fh,Lednicky:1981su}
\be
\label{def-fun-cor}
\mathcal{R}({\bf p}_p, {\bf p}_{^3{\rm He}})
=\int d^3 {\bf r}_p \, d^3 {\bf r}_{^3{\rm He}}\, 
D({\bf r}_p) \, D({\bf r}_{^3{\rm He}}) \,
|\psi_{p\, ^3{\rm He}}({\bf r}_{\rm p},{\bf r}_{^3{\rm He}})|^2,
\ee
where $\psi_{p\, {^3{\rm He}}}({\bf r}_p,  {\bf r}_{^3{\rm He}})$ is the wave function of the $p\!-\!^3{\rm He}$ pair and $D({\bf r}_i), i= p,{^3{\rm He}}$ is the normalized source function which is  assumed further on to be of the Gaussian form (\ref{Gauss-source}).

The correlation function (\ref{def-fun-cor}), as the formation rate (\ref{A-form-rate}), is written as for the instantaneous emission of the two particles but the time duration of the emission process can be easily incorporated \cite{Koonin:1977fh}. In case of an isotropic Gaussian source function, the time duration $\tau$, as already mentioned, enlarges the effective radius of the source from $R_s$ to $\sqrt{R_s^2 + v^2\tau^2}$ where $v$ is the velocity of the particle pair relative to the source. 

Similarly to the formation process and for the same reasons we consider the $p\!-\!^3{\rm He}$ correlations in the center-of-mass frame of the pair and we treat the formula (\ref{def-fun-cor}) as nonrelativistic even so the nuclide and proton momenta can be  relativistic in both the rest frame of the source and in the laboratory frame. 

We introduce the center-of-mass variables
\be
\label{zmien-sr-masy}
\left\{ \begin{array}{ll}
{\bf R}=\frac{{\bf r}_p+3{\bf r}_{^3{\rm He}}}{4} ,
\\[2mm]
{\bf r} = {\bf r}_p -{\bf r}_{^3{\rm He}} ,
\end{array} \right.
~~~~~~~~~~~~~~~~~~~~~~
\left\{ \begin{array}{ll}
{\bf r}_p ={\bf R}+\frac{3{\bf r}}{4} ,
\\[2mm]
{\bf r}_{^3{\rm He}}={\bf R}-\frac{\bf r}{4} ,
\end{array} \right.
\ee
and write down the wave function as 
\be
\psi_{p\, {^3{\rm He}}}({\bf r}_p, {\bf r}_{^3{\rm He}}) 
= e^{i{\bf P}{\bf R}} \, \phi_{\bf q}({\bf r}) ,
\ee 
where $\phi_{\bf q}({\bf r})$ is the wave function of relative motion of $p$ and $^3{\rm He}$ and ${\bf q} \equiv \frac{1}{4}(3{\bf p}_p - {\bf p}_{^3{\rm He}})$ is the proton momentum in the center-of-mass frame of the pair. The correlation function then equals
\be
\label{Rcz}
\mathcal{R}({\bf q})=\int d^3 {\bf r}\, d^3 {\bf R} \, 
D\Big({\bf R}+\frac{3{\bf r}}{4}\Big)\,D\Big({\bf R}-\frac{\bf r}{4}\Big) \,
|\phi_{\bf q} ({\bf r})|^2 .
\ee
Defining the `relative' source
\be
\label{D_wz}
D_r({\bf r}) \equiv \int d^3{\bf R}\;D\Big({\bf R}+\frac{3{\bf r}}{4}\Big) \:
D\Big({\bf R}-\frac{\bf r}{4}\Big) ,
\ee
which for the Gaussian parametrization (\ref{Gauss-source}) equals 
\be
\label{Gauss-source-r}
D_r({\bf r})= \frac{1}{\big(4\pi R_s^2\big)^{3/2}} \; e^{-\frac{{\bf r}^2}{4R_s^2}},
\ee
the correlation function acquires the well-known form
\be
\label{fun-corr-el}
\mathcal{R}({\bf q})=\int d^3 {\bf r} \: D_r({\bf r})|\phi_{\bf q}({\bf r})|^2.
\ee
We note that once the single particle source (\ref{Gauss-source}) is independent of particle's mass, the source function (\ref{Gauss-source-r}) is also mass independent even so the transformation to the center-of-mass variables (\ref{zmien-sr-masy}) depends on particle masses. 

\subsection{$^3{\rm He}$ treated as a bound state}

Taking into account that the nucleus $^3{\rm He}$ is a bound state of $(p,p,n)$ formed due to final state interactions at the same time when the correlation among ${^3{\rm He}}$ and $p$ is generated, the correlation function is defined as
\be
\label{cross_section}
\frac{d P_{p \,{^3{\rm He}}}}{d^3 p_p d^3 p_{^3{\rm He}}} 
= \mathcal{R}({\bf p}_p, {\bf p}_{^3{\rm He}})\,\mathcal{A}_3\,
\frac{d P_p}{d^3 p_p}
\frac{d P_N}{d^3(p_{^3{\rm He}}/3)}
\frac{d P_N}{d^3(p_{^3{\rm He}}/3)}
\frac{d P_N}{d^3(p_{^3{\rm He}}/3)},
\ee
where $\mathcal{A}_3$ is the formation rate of a nucleus $^3{\rm He}$ defined through the relation (\ref{A-mom-dis}) and  given by the formula (\ref{A-form-rate}) both for $A=3$. The product of the formation rate and correlation function is
\ba
\label{RA}
\mathcal{R}({\bf p}_{\rm p}, {\bf p}_{^3{\rm He}})\mathcal{A}_3 &=&
g_S g_I (2\pi)^6 \int d^3 {\bf r}_{\rm p} d^3 {\bf r}_{1}d^3 {\bf r}_{2}d^3 {\bf r}_{3} \, 
D({\bf r}_{\rm p})D({\bf r}_1)D({\bf r}_2)D({\bf r}_3)
|\psi_{{\rm p}\, ^3{\rm He}}({\bf r}_{\rm p},{\bf r}_1,{\bf r}_2, {\bf r}_3)|^2,
\ea
where again $D({\bf r}_i)$ with $i= p,1, 2, 3$ is the source function while $\psi_{{\rm p}\, ^3{\rm He}}({\bf r}_{\rm p},{\bf r}_1,{\bf r}_2, {\bf r}_3)$ is the wave function of  $p$ and ${^3{\rm He}}$. 

\subsubsection{Formation rate}

Let us first compute the formation rate of $^3{\rm He}$. Using the Jacobi variables for a system of three particles with equal masses which are
\be 
\label{Jacobi-3}
\left\{ \begin{array}{ll}
{\bf R} \equiv \frac{1}{3}({\bf r}_1+{\bf r}_2+{\bf r}_3) ,
\\[2mm]
{\bf x} \equiv {\bf r}_2-{\bf r}_1,
\\[2mm]
{\bf y} \equiv {\bf r}_3-\frac{1}{2}({\bf r}_1+{\bf r}_2) ,
\end{array} \right.
~~~~~~~~~~~~~~~~~~
\left\{ \begin{array}{ll}
{\bf r}_1 = {\bf R} -\frac{1}{2} {\bf x} - \frac{1}{3} {\bf y} ,
\\[2mm]
{\bf r}_2 = {\bf R} + \frac{1}{2}{\bf x} - \frac{1}{3} {\bf y} ,
\\[2mm]
{\bf r}_3 = {\bf R} + \frac{2}{3} {\bf y} ,
\end{array} \right.
\ee
and writing down the wave function as 
\be
\psi_{^3{\rm He}}({\bf r}_1, {\bf r}_2, {\bf r}_3) = e^{i{\bf P}{\bf R}} \, \phi_{^3{\rm He}}({\bf x}, {\bf y}),
\ee
with $\phi_{^3{\rm He}}({\bf x}, {\bf y})$ being the wave function of relative motion, the formation rate (\ref{A-form-rate}) for $A=3$ equals
\be
\label{A3}
\mathcal{A}_3 = g_S g_I (2\pi)^6 \int d^3 {\bf x}\,  d^3{\bf y}\,
 D_r({\bf x},{\bf y}) \, |\phi_{^3{\rm He}}({\bf x}, {\bf y})|^2,
\ee
where
\be
\label{Dr-x-y}
D_r({\bf x},{\bf y}) \equiv \int d^3R \, 
D\Big({\bf R} -\frac{1}{2} {\bf x} - \frac{1}{3} {\bf y}\Big) \,
D\Big({\bf R} +\frac{1}{2} {\bf x} - \frac{1}{3} {\bf y}\Big) \,
D\Big({\bf R} + \frac{2}{3} {\bf y}\Big) 
= \frac{1}{\big(2\sqrt{3}\pi R_s^2\big)^3}\, 
e^{-\frac{{\bf x}^2}{4R_s^2} -\frac{{\bf y}^2}{3R_s^2}} .
\ee
The latter equality holds for the Gaussian parametrization (\ref{Gauss-source}). We note that the relative source (\ref{Dr-x-y}) is normalized that is
\be
\int d^3{\bf x}\, d^3{\bf y} \, D_r({\bf x},{\bf y}) = 1 .
\ee

Let us derive an explicit formula of the formation rate of $^3{\rm He}$ that will be needed later on. Using the source function (\ref{Dr-x-y}) and choosing the wave function of $^3{\rm He}$ in the Gaussian form such that
\be
|\phi_{^3{\rm He}}({\bf x}, {\bf y})|^2 =  \Big(\frac{\beta}{\pi}\Big)^3
e^{- \beta \big( ({\bf r}_1 - {\bf r}_2)^2 + ({\bf r}_2 - {\bf r}_3)^2 + ({\bf r}_1 - {\bf r}_3)^2 \big)}
= 3^{3/2} \Big(\frac{\beta}{\pi}\Big)^3 
e^{- \beta ( \frac{3}{2} {\bf x}^2 + 2{\bf y}^2)} ,
\ee
where the root-mean-square radius of $^3{\rm He}$ equals $R_3 =1/\sqrt{3 \beta}$, the formation rate (\ref{A3}) is
\be
\label{A3-Gauss}
\mathcal{A}_3 = \frac{\pi^3}{3^{3/2}}  \frac{1} {(R_s^2 + \frac{1}{2} R_3^2)^3 } .
\ee
The spin and isospin factors, which are $g_S=1/2$ and $g_I = 1/4$, are included here.
\subsubsection{Correlation function}
To compute the  $p\!-\!^3{\rm He}$ correlation function we use the Jacobi variables for a system of four particles defined by Eqs.~(\ref{Jacobi-4}) and we write down the wave function as 
\be
\label{wave-function-p-3He}
\psi_{{\rm p}\, ^3{\rm He}}({\bf r}_{\rm p},{\bf r}_1,{\bf r}_2, {\bf r}_3) 
= e^{i{\bf P}{\bf R}} \, \varphi_{\bf q} ({\bf z}) \, \phi_{^3{\rm He}}({\bf x}, {\bf y}),
\ee
where $\varphi_{\bf q} ({\bf z})$ is the wave function of relative motion in the center of mass of  $p\!-\!^3{\rm He}$ system. The integral expression (\ref{RA}) then equals
\ba
\label{RA-1}
\mathcal{R}({\bf q}) \, \mathcal{A}_3 
&=&  g_S g_I (2\pi)^6 \int d^3 {\bf R} \, d^3 {\bf x} \, d^3 {\bf y} \, d^3 {\bf z} \, 
D\Big({\bf R} -\frac{1}{2} {\bf x} - \frac{1}{3} {\bf y} - \frac{1}{4} {\bf z}\Big) \,
D\Big({\bf R} +\frac{1}{2} {\bf x} - \frac{1}{3} {\bf y}- \frac{1}{4} {\bf z}\Big) 
\\[2mm] \nn 
&&~~~~~~~~~~~~~~~~~~~~~~~~~~~~~~
\times
D\Big({\bf R} + \frac{2}{3} {\bf y}- \frac{1}{4} {\bf z}\Big) \,
D\Big({\bf R} + \frac{3}{4} {\bf z}\Big) \, 
|\varphi_{\bf q} ({\bf z})|^2 \,|\phi_{^3{\rm He}}({\bf x}, {\bf y})|^2 .
\ea

Further on calculations are performed using the Gaussian parametrization (\ref{Gauss-source}). Since the Jacobi variables have the property (\ref{Jacobi-property-1}), the elementary integral over ${\bf R}$ factors out and the formula (\ref{RA-1}) equals
\be
\label{RA-3}
\mathcal{R}({\bf q}) \, \mathcal{A}_3
= g_S g_I (2\pi)^6 \int  d^3 {\bf x}\, d^3 {\bf y}\, d^3 {\bf z}\, 
D_r({\bf x},{\bf y}) \, D_{4r}({\bf z}) \,
 |\varphi_{\bf q} ({\bf z})|^2 \,|\phi_{^3{\rm He}}({\bf x}, {\bf y})|^2 ,
\ee
where $D_r({\bf x},{\bf y})$ is, as previously, given by Eq.~(\ref{Dr-x-y}) and 
\be
\label{Dz}
 D_{4r}({\bf z}) = \Big(\frac{3}{8\pi R_s^2}\Big)^{3/2} e^{-\frac{3 {\bf z}^2}{8R_s^2}} ,
\ee
which is also normalized to unity. Since one recognizes the formation rate (\ref{A3}) in the right-hand-side of Eq.~(\ref{RA-3}), the rate $ \mathcal{A}_3$ factors out and the correlation function simplifies to the form
\be
\label{fun-cor-bound}
\mathcal{R}({\bf q})=\int d^3 {\bf r}\, D_{4r}({\bf r}) \,|\varphi_{\bf q}({\bf r})|^2 ,
\ee
where we have changed the integral variable ${\bf z}$ into ${\bf r}$. 

The formulas (\ref{fun-cor-bound}) and (\ref{fun-corr-el}) are almost the same but the source functions differ. The source radius of nuclei $^3{\rm He}$ treated as bound states is bigger by the factor $\sqrt{3/2}\approx 1.22$ than that of  the `elementary' nuclides $^3{\rm He}$. When the source radius inferred from the $p\!-\!p$ correlation function is the same as the radius obtained from the $p\!-\!^3{\rm He}$ correlation function, it means that the nuclides $^3{\rm He}$ are directly emitted from the fireball. If the radius is bigger by $\sqrt{3/2}$, the nuclides are formed due to final state interactions. The question is, however, whether the $p\!-\!^3{\rm He}$ correlation function is sensitive enough to the change of source radius from $R_s$ to $\sqrt{3/2}\,R_s$. The question is discussed in the next section.

\section{$p\!-\!^3{\rm He}$ correlation function}
\label{sec-corr-fun-comp}

To compute the correlation function (\ref{fun-corr-el}) or (\ref{fun-cor-bound}) we have to specify the wave function $\varphi_{\bf q}({\bf r})$. Following Lednick\'y and Luboshitz \cite{Lednicky:1981su}, we choose the function in the asymptotic scattering form that is 
\be
\label{wave-fun-asym-scatt}
\varphi_{\bf q}({\bf r})=e^{i{\bf q}{\bf r}}+f(q,\theta)\frac{e^{iqr}}{r},
\ee
where the amplitude $f(q,\theta)$ depends in general on the momentum $q$ and scattering angle $\theta$.  The correlation function (\ref{fun-cor-bound}), which assumes that the nuclides $^3{\rm He}$ are formed after the nucleons are emitted from the source, is then expressed as
\ba
\label{fun-corr-gen}
\mathcal{R}({\bf q}) 
= 1 + \Big(\frac{3}{8\pi R_s^2}\Big)^{3/2}
\int d^3 {\bf r} \, e^{-\frac{3{\bf r}^2}{8R_s^2}} 
\Bigg(\frac{|f(q,\theta)|^2}{r^2} +2\Re\bigg(f(q,\theta)\frac{e^{i(qr-{\bf q}{\bf r})}}{r}\bigg) \Bigg) ,
\ea
where the source function (\ref{Dz}) is used. The expression analogous to Eq.~(\ref{fun-corr-gen}) corresponding to the correlation function (\ref{fun-corr-el}), which assumes that the nuclides $^3{\rm He}$ are directly emitted from the source, can be obtained from the formula (\ref{fun-corr-gen}) by means of the replacement $R_s \to \sqrt{3/2}\,R_s$.

\begin{figure}[t]
\begin{minipage}{87mm}
\centering
\includegraphics[scale=0.28]{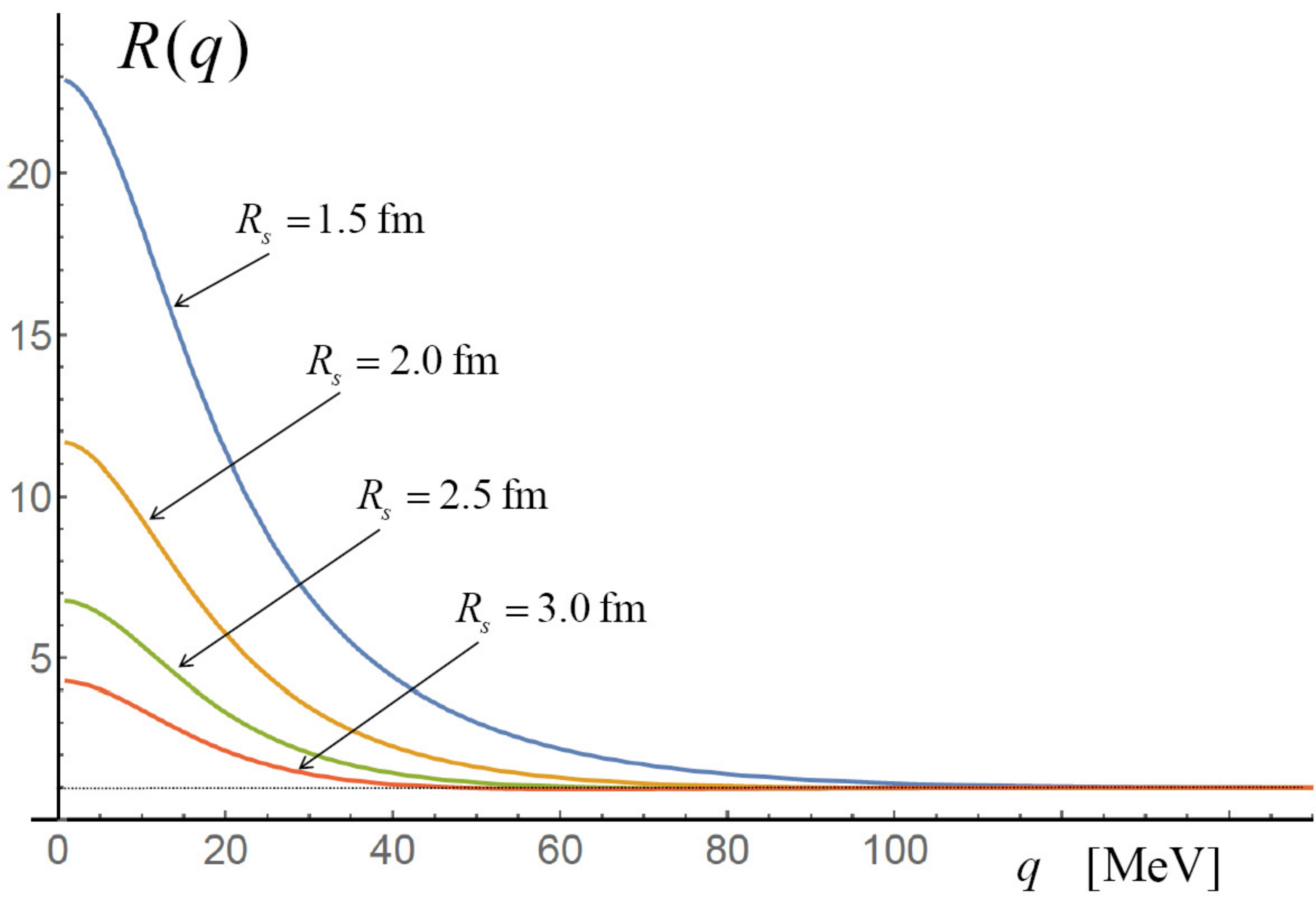}
\vspace{-6mm}
\caption{The spin-average $p\!-\!^3{\rm He}$ correlation function which takes into account only the $s-$wave scattering.}
\label{Fig-s-scatter}
\end{minipage}
\hspace{1mm}
\begin{minipage}{87mm}
\centering
\vspace{5mm}
\includegraphics[scale=0.28]{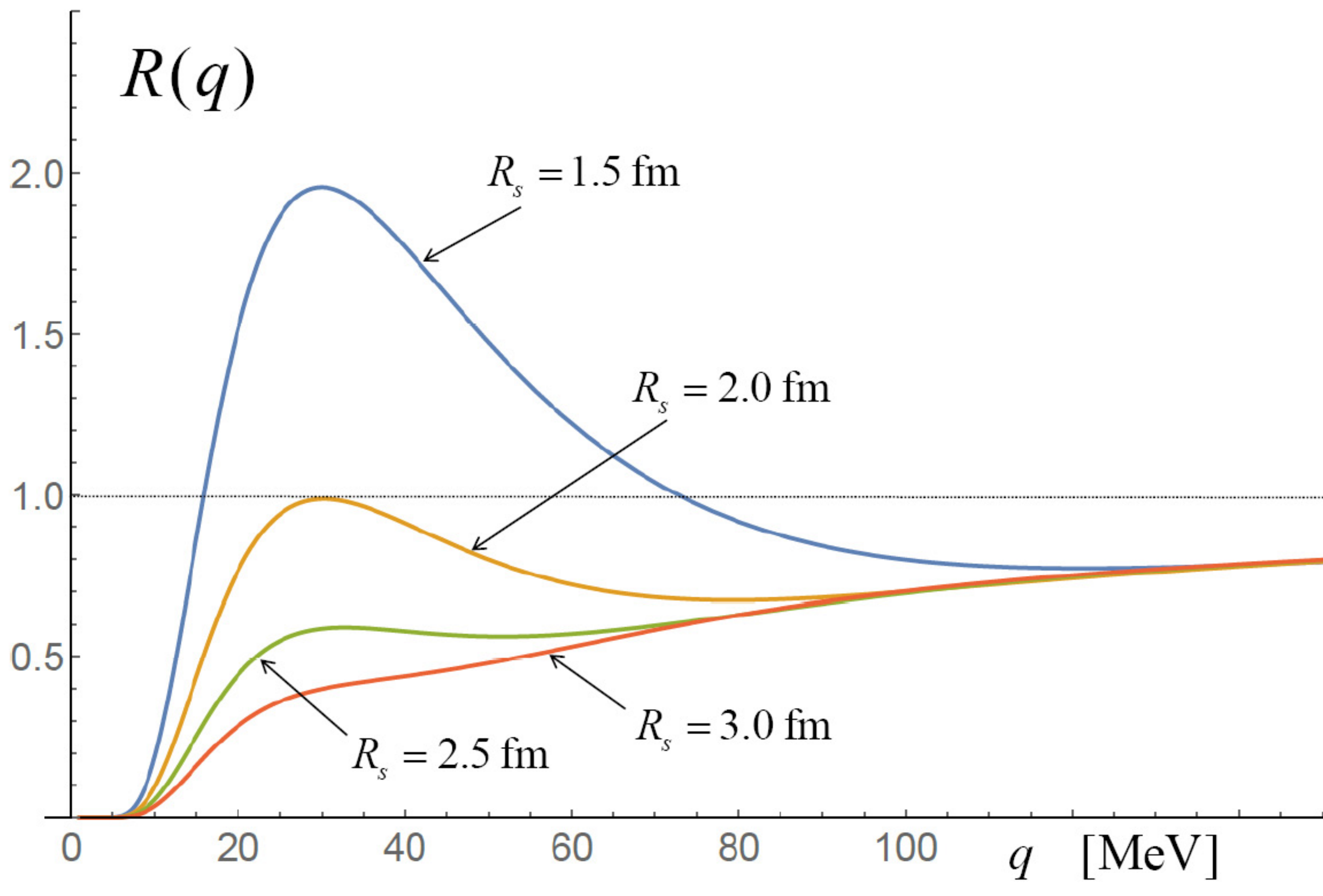}
\vspace{-1mm}
\caption{The spin-average $p\!-\!^3{\rm He}$ correlation function which takes into account the $s-$wave scattering and Coulomb repulsion.}
\label{Fig-s-scatter-Gamow}
\end{minipage}
\hspace{2mm}
\end{figure}

\subsection{Scattering in $s-$wave}

The correlation function significantly differs from unity only in the domain of small momenta $q$, and thus the amplitude $f(q,\theta)$ can be approximated by its $s-$wave contribution $f_0(q)$ which is isotropic and thus the amplitude is independent of $\theta$.  The angular integral in Eq.~(\ref{fun-corr-gen}) becomes elementary and the correlation function thus equals
\be
\label{fun-cor-2}
\mathcal{R}_0({\bf q})= 1 + \frac{3}{4 R_s^2} |f_0(q)|^2 
- \frac{3}{4 R_s^2 q } \bigg( 1 -  e^{-\frac{8 q^2 R_s^2}{3}}\bigg)  \Im f_0(q)
+ \frac{3^{3/2}}{2^{5/2}\pi^{1/2}R_s^3 q} \,\Re f_0(q) 
\int_0^{\infty}dr \, e^{-\frac{3r^2}{8R_s^2}}\,\sin(2qr) .
\ee
The remaining integral in Eq.~(\ref{fun-cor-2}) needs to be taken numerically. 

Since both proton and $^3{\rm He}$ have spin 1/2, there is a singlet (spin zero) and a triplet (spin one) channel of the $p\!-\!^3{\rm He}$ scattering. The corresponding scattering lengths are sizable \cite{Daniels:2010af}, see also \cite{Kirscher:2011uc}, and are equal to
\be
a_s=11.1\; {\rm fm}, 
~~~~~~~~~~
a_t=9.05\;{\rm fm}.
\ee
According to the analysis \cite{Levashev:2008ec}, the effective ranges in both channels do not exceed 2 fm and are significantly smaller than the corresponding scattering lengths. Therefore, the effective ranges can be ignored and the $s-$wave amplitude is written as
\be
\label{f^0}
f_0^{s,t}(q)=\frac{-a_{s,t}}{1+iq\,a_{s,t}} .
\ee
The correlation function (\ref{fun-cor-2}) of the singlet and triplet channel thus equals
\be
\label{Rx}
\mathcal{R}_0^{s,t}({\bf q})=1+ \frac{3}{4 R_s^2}\, \frac{a_{s,t}^2}{1+q^2a_{s,t}^2}e^{-\frac{8q^2R_s^2}{3}}
- \bigg(\frac{3^3}{2^5\pi }\bigg)^{1/2}\frac{a}{q R_s^3(1+q^2a_{s,t}^2)} 
\int_0^{\infty}dr \, e^{-\frac{3 r^2}{8R_s^2}}\sin(2qr) .
\ee

Protons and ${^3{\rm He}}$ nuclei, which are emitted from a fireball, are assumed to be unpolarized, and consequently the spin-average correlation function is 
\be
\mathcal{R}({\bf q})=\frac{1}{4}\mathcal{R}_0^s({\bf q})+\frac{3}{4}\mathcal{R}_0^t({\bf q}),
\ee
where the weight factors $1/4$ and $3/4$ reflect the numbers of singlet and triplet states. In Fig.~\ref{Fig-s-scatter} we show the spin-average $p\!-\!^3{\rm He}$ correlation function which takes into account only the $s-$wave scattering. There is a strong positive correlation due to the attractive interaction of $^3{\rm He}$ and $p$.

\subsection{Coulomb effects}

When one deals with charged particles, the formula (\ref{wave-fun-asym-scatt}) needs to be modified as the long-range electrostatic interaction influences both the incoming and outgoing waves, see the formula (135.8) of the textbook \cite{Landau-Lifshitz-1988}. However, the Coulomb effect can be approximately taken into account \cite{Gmitro:1986ay} by multiplying the correlation function by the Gamow factor which for repelling particles  equals
\be 
\label{Gamow}
G(q) = {2 \pi \over a_B q} \,
{1 \over {\rm exp}\big({2 \pi \over a_B q}\big) - 1} ,
\ee
where $a_B = 1/(2\mu \alpha)$ is the Bohr radius of the ${^3{\rm He}\!-\!p}$ system with $\mu=703.3$~MeV and $\alpha = 1/137$ being the reduced mass and the fine structure constant. The factor of 2 takes into account the double charge of ${^3{\rm He}}$. 

In Fig.~\ref{Fig-s-scatter-Gamow} we show the spin-average $p\!-\!^3{\rm He}$ correlation function which takes into account the $s-$wave scattering and Coulomb repulsion. As seen, the electrostatic interaction strongly modifies the correlation. 

\begin{figure}[t]
\begin{minipage}{87mm}
\centering
\includegraphics[scale=0.27]{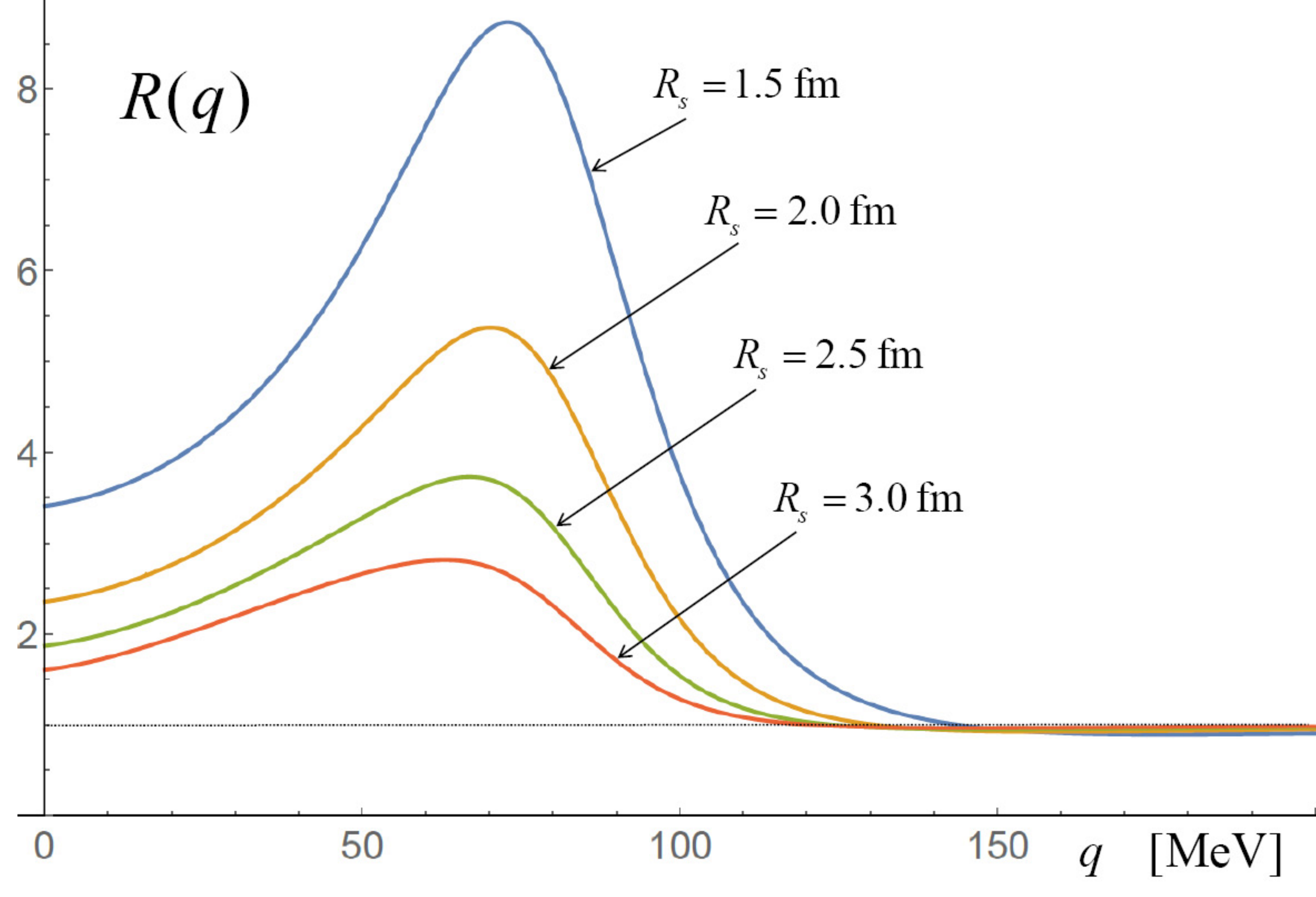}
\vspace{-2mm}
\caption{The $p\!-\!^3{\rm He}$ correlation function which takes into account only the resonance of $^4{\rm Li}$.}
\label{Fig-resonance-1}
\end{minipage}
\hspace{1mm}
\begin{minipage}{87mm}
\centering
\vspace{4mm}
\includegraphics[scale=0.27]{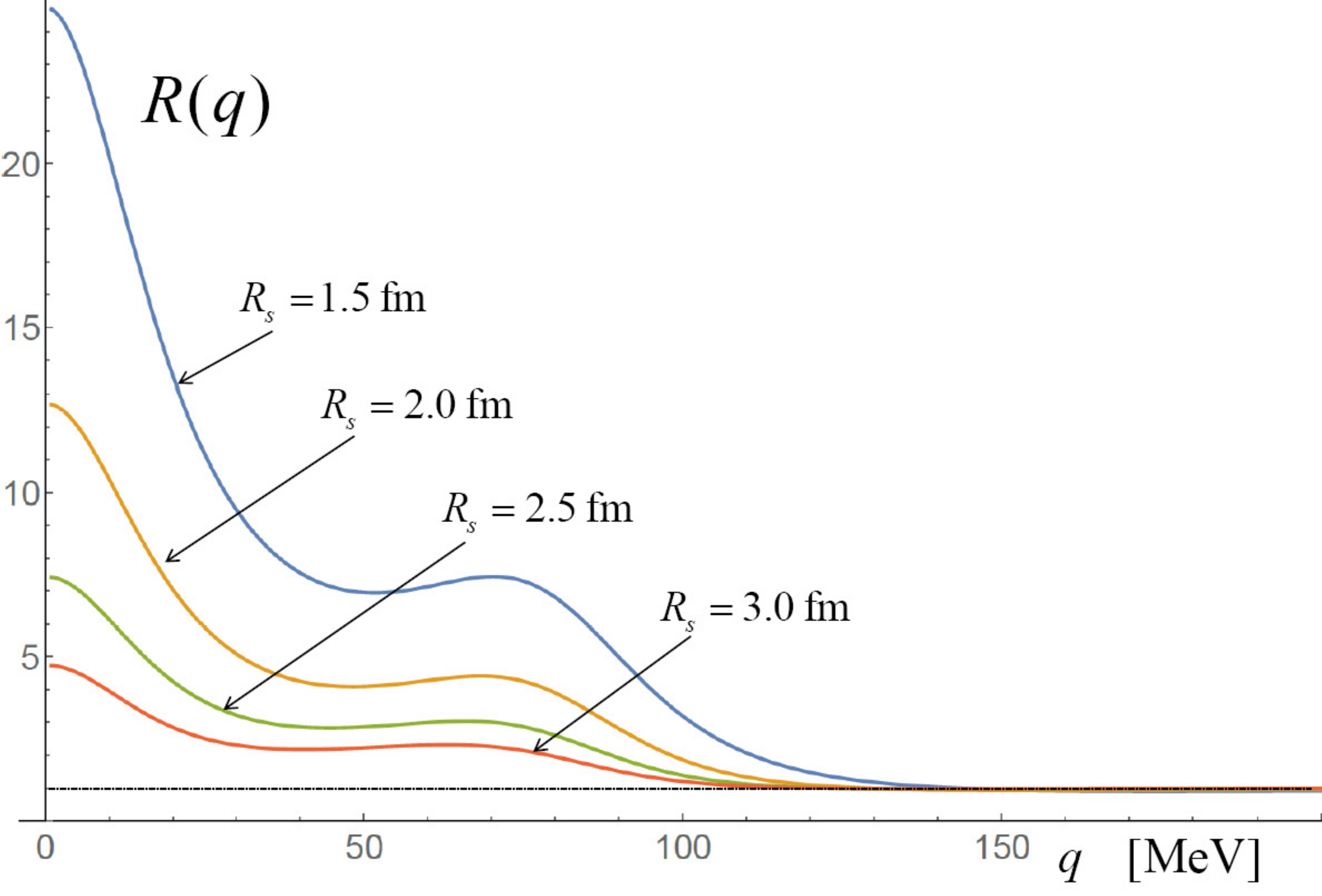}
\vspace{-2mm}
\caption{The spin-average $p\!-\!^3{\rm He}$ correlation function which takes into account  the $s-$wave scattering and the resonance of $^4{\rm Li}$ in the triplet channel.}
\label{Fig-resonance-1-s-wave}
\end{minipage}
\hspace{2mm}
\end{figure}

\subsection{Resonance interaction}

The nuclide $^4{\rm Li}$, which presumably has a cluster structure of $p\!-\!^3{\rm He}$, manifests itself as a resonances in the $p\!-\!^3{\rm He}$ scattering. The resonance mass equals $\Delta_E = 4,07$ MeV above the sum of masses of proton and ${^3{\rm He}}$ and its width is $\Gamma = 6,03$ MeV \cite{Tilley:1992zz}. 

The amplitude, which takes into account the resonance scattering, is (see the formula (134.12) of the textbook \cite{Landau-Lifshitz-1988})
\be
f(q,\theta)=f_0(q) + f_l^r(q) \, P_l(\cos\theta),
\ee
where $f_0(q)$ is the $s-$wave scattering amplitude,  and $f_l^r(q)$ is the resonance contribution with $l$ being the orbital momentum of the resonance state, $P_l(\cos\theta)$ is the Legendre polynomial which for $l=1$ equals $P_1(\cos\theta) = \cos\theta$. 

The resonance amplitude is of the Breit-Wigner form
\be
\label{B-W-amplitude}
f_l^r(q) \equiv - \lambda_R\ \frac{2l+1}{q_0}\frac{\frac{1}{2}\Gamma}{E-E_0+\frac{1}{2}i\Gamma},
\ee
where $E_0$ and $\Gamma$ are the resonance energy and its width and the parameter $\lambda_R$, which is assumed to be real, controls a strength of the resonance. In our numerical calculations we put $\lambda_R=1$ but, as we discuss further on, the parameter $\lambda_R$ can be and should be inferred from experimental data. 

\begin{figure}[t]
\begin{minipage}{87mm}
\centering
\includegraphics[scale=0.27]{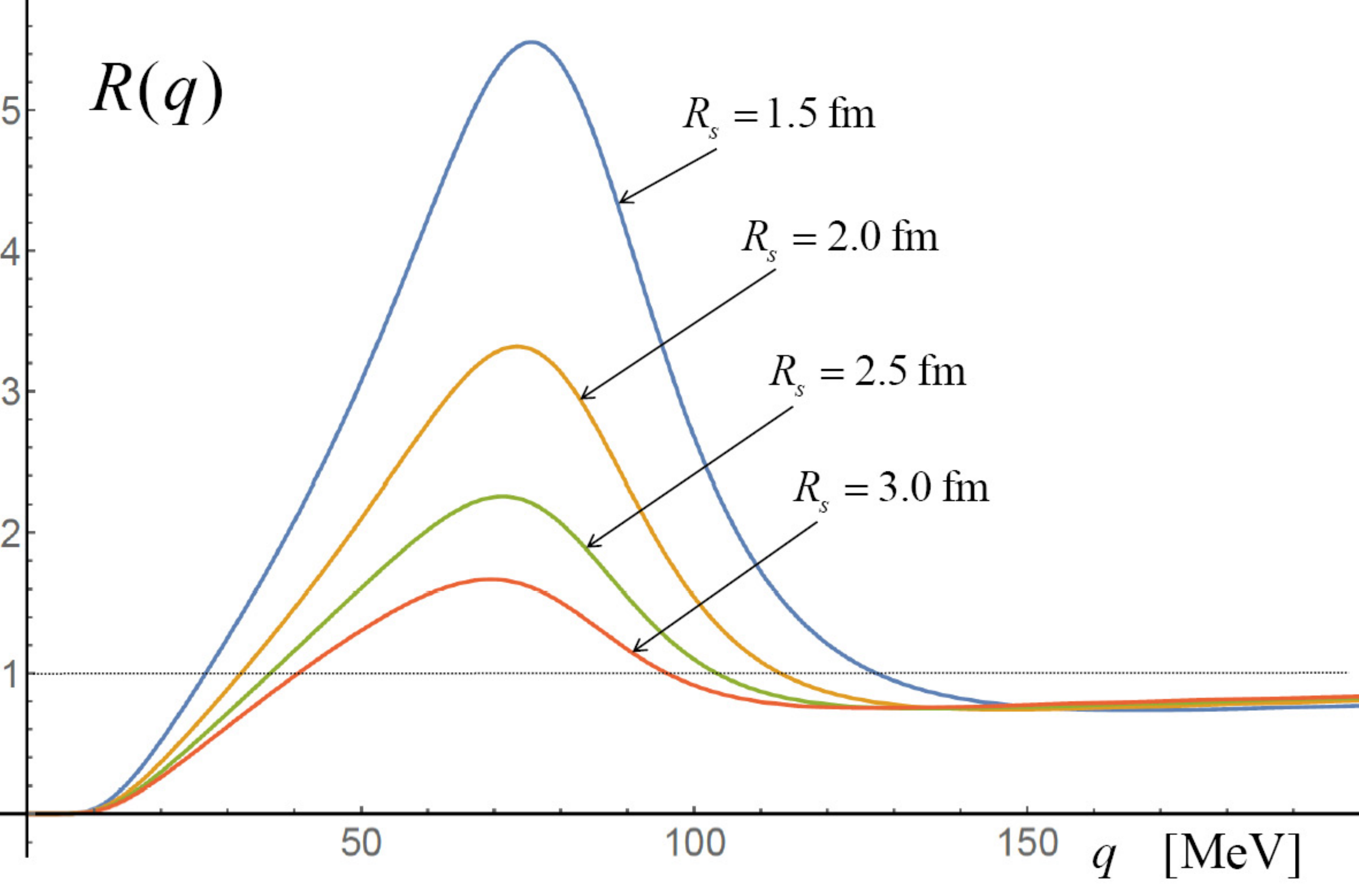}
\vspace{-2mm}
\caption{The $p\!-\!^3{\rm He}$ correlation function which takes into account the resonance $^4{\rm Li}$ and the Coulomb repulsion.}
\label{Fig-resonance-1-Coulomb}
\end{minipage}
\hspace{1mm}
\begin{minipage}{87mm}
\centering
\vspace{3mm}
\includegraphics[scale=0.27]{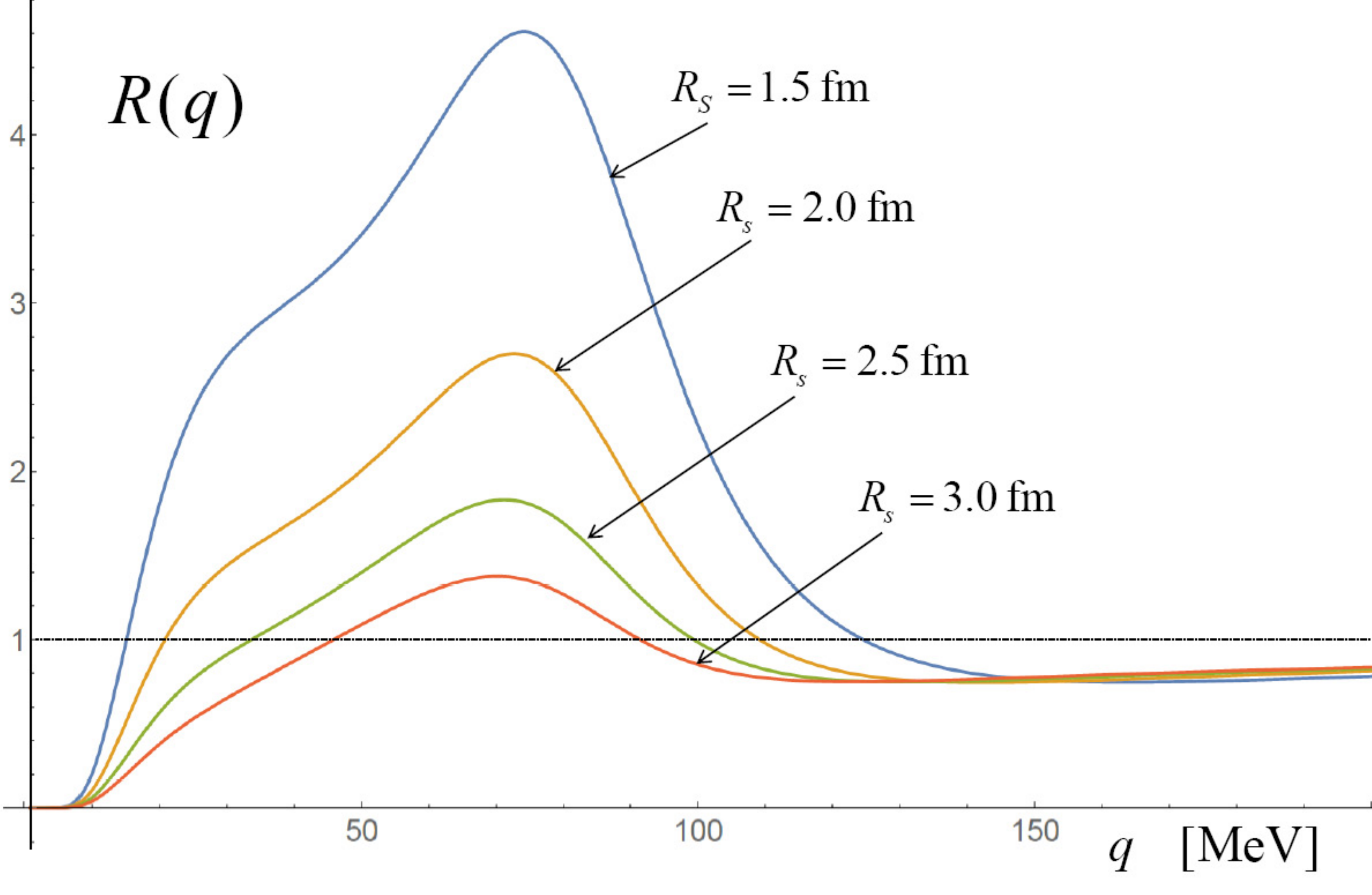}
\vspace{-2mm}
\caption{The spin-average $p\!-\!^3{\rm He}$ correlation function which takes into account the $s-$wave scattering, the resonance $^4{\rm Li}$ in the triplet channel and the Coulomb repulsion.}
\label{Fig-resonance-1-s-wave-Coulomb}
\end{minipage}
\hspace{2mm}
\end{figure}

When compared to the original formula (134.12) from the textbook \cite{Landau-Lifshitz-1988}, we have introduced the parameter $\lambda_R$ and we have replaced the factor $(2l+1)/q$ by $(2l+1)/q_0$, where $q_0$ corresponds to the energy $E_0$, to avoid the divergence of the amplitude at $q=0$. The modification is legitimate as, strictly speaking, the amplitude is valid only in the vicinity of the resonance. However, it would be more appropriate to include a momentum dependence of the resonance width. Actually, our prescription to regulate the divergence is close to the assumption that the momentum-dependent width equals $\Gamma(q) = q \Gamma_0/q_0$ where $\Gamma_0$ is the width in the vicinity of the peak. The width of a resonance of orbital angular momentum $l=1$ behaves as $\Gamma(q) \sim q^3$ when $q \rightarrow 0$, see the formula (10.58) from the textbook \cite{Sachs-1955}, and the divergence at $q=0$ is again removed. However, it is unclear how to parameterize the width of $^4{\rm Li}$ in a broader domain of $q$ because experimental information on the $^4{\rm Li}$ resonance is rather scarce. Nevertheless, it is important to realize that the $p\!-\!^3{\rm He}$ correlation function is heavily dominated by the Coulomb repulsion in the domain of $q \lesssim 20$ MeV, see Fig.~\ref{Fig-resonance-1-Coulomb}. Therefore, it does not much matter how the resonance contribution is parameterized in this domain. 

In case of the  $^4{\rm Li}$ resonance, the energy difference, which enters the amplitude, is 
\be
E-E_0 = \frac{q^2}{2\mu} -  \Delta_E ,
\ee
and the momentum $q_0$, which corresponds to the resonance peak, is $q_0=75.7$~MeV. 

The correlation function, which takes into account the resonance interaction, is found as
\ba
\label{fun-corr-gen-final}
\mathcal{R}({\bf q}) 
= \mathcal{R}_0({\bf q}) + \Big(\frac{3}{8\pi R_s^2}\Big)^{3/2} |f_l^r(q)|^2 J_l 
+ 2 \Big(\frac{3}{8\pi R_s^2}\Big)^{3/2}
\Big(\Re f_l^r(q) \, \Re K_l(q) -  \Im f_l^r(q)  \, \Im K_l(q) \Big) ,
\ea
where $\mathcal{R}_0({\bf q})$ is given by Eq.~(\ref{fun-cor-2}). For $l=1$ the coefficient $J_l$ and the function $K_l(q)$ are
\be
J_1 = \frac{2^{5/2} \pi^{3/2}}{3^{3/2}} \, R_s, 
\ee
\ba
\Re K_1(q) &=& - \frac{2\pi}{q} \int_0^\infty dr \, e^{-\frac{3{\bf r}^2}{8R_s^2}} \, \sin(2qr) 
+ \frac{4\pi}{q^2} \int_0^\infty dr \, e^{-\frac{3{\bf r}^2}{8R_s^2}} \;  
 \frac{\sin^2(qr)}{r}  ,
\\[2mm] 
\label{ImK1}
\Im K_1(q) &=& \frac{2^{3/2} \pi^{3/2} R_s}{3^{1/2}q} \Big( 1 + e^{- \frac{8 q^2 R_s^2}{3}} \Big) 
- \frac{2\pi}{q^2} \int_0^\infty dr \, e^{-\frac{3{\bf r}^2}{8R_s^2}} \;  
\frac{\sin(2qr)}{r} .
\ea
A computation of $\Im K_1(q)$ requires some care as both terms in Eq.~(\ref{ImK1}) diverge when $q \to 0$. However, the divergences cancel out and $\Im K_1(q)$ vanishes for $q = 0$. 

In Fig.~\ref{Fig-resonance-1} we show the $p\!-\!^3{\rm He}$ correlation function which takes into account only the resonance interaction. The peak at $q=q_0=75.7$~MeV is well seen. Fig.~\ref{Fig-resonance-1-s-wave} presents the spin-average correlation function which includes the $s-$wave scattering and the resonance $^4{\rm Li}$. The resonance contributes only to the triplet correlation function and the spin-average correlation function is obtained summing up the singlet and triplet functions with the weights $1/4$ and $3/4$, respectively. The resonance correlation function which additionally takes into account the Coulomb repulsion is shown in Fig.~\ref{Fig-resonance-1-Coulomb}. Finally, we show in Fig.~\ref{Fig-resonance-1-s-wave-Coulomb} the spin-average correlation function which takes into account the $s-$wave scattering, the resonance in the triplet channel and the Coulomb repulsion. One observes that the  $s-$wave scattering and Coulomb repulsion strongly deform the resonance peak. 

As already mentioned, the $p\!-\!^3{\rm He}$ correlation function was measured in $^{40}{\rm Ar}-$induced reactions on $^{197}{\rm Au}$ at the collision energy per nucleon of 60 MeV \cite{Pochodzalla:1987zz}. The shape of the $p\!-\!^3{\rm He}$ correlation function with the peak of $^4{\rm Li}$ well seen, which is shown in Fig.~6 of Ref. \cite{Pochodzalla:1987zz}, is very similar to that presented in our Fig.~\ref{Fig-resonance-1-s-wave-Coulomb} for $R_s = 3.0$ fm. However, a quantitative comparison is not possible because the measurement is not very precise and its details are not given. 

As noted in the Introduction, the  calculation of the $p\!-\!^3{\rm He}$ correlation function has been recently presented in \cite{Xi:2019vev}. The correlation function, which includes the $s-$wave scattering and Coulomb repulsion, has been obtained, as we have done, using the method by Lednick\'y and Luboshitz \cite{Lednicky:1981su}. The resonance interaction, however, has not been taken into account in the correlation function but the $p\!-\!^3{\rm He}$ pairs coming from the two-body decays of $^4{\rm Li}$ have been generated by means of a Monte Carlo method and added to the correlation function. Therefore, there is no interference of the incoming and outgoing waves modified by the resonance interaction and the resonance contribution to the correlation function does not depend on the source radius. 

Since the authors of Ref.~\cite{Xi:2019vev} used the ``spherically symmetric Gaussian distribution source with a radius of 5.5 fm", which we identify with the RMS radius, their results presented in Figs. 2 and 3 are directly comparable to the $p\!-\!^3{\rm He}$ correlation function shown in our Fig.~\ref{Fig-resonance-1-s-wave-Coulomb} for $R_s = 3$ fm, which corresponds to the RMS equal $\sqrt{3} R_s = 5.2$ fm. The correlation functions evidently differ, the general shape is different. The width of the resonance peak of Ref.~\cite{Xi:2019vev} is less than 10 MeV while ours is a few tens of MeV. As already mentioned, our correlation function is very similar to the measured one \cite{Pochodzalla:1987zz}.

A measurement of the $p\!-\!^3{\rm He}$ correlation function in relativistic heavy-ion collisions at LHC is difficult but feasible \cite{private}. The main problem is to collect a sufficient statistics. The number of correlated $p\!-\!^3{\rm He}$ pairs is of the same order as that of $^4{\rm He}$ nuclides which are registered at midrapidity roughly one per million central Pb-Pb collisions at LHC. Therefore, millions of central events are needed to construct the correlation function. 

\section{Yield of $^4{\rm Li}$}
\label{sec-yield}

As discussed in the previous section, the resonance peak of the correlation function is distorted by the Coulomb repulsion and $s-$wave scattering. So, it is not evident how to infer the resonance yield from the distribution of the $p\!-\!^3{\rm He}$  pairs.  

To derive the formula, which gives the yield of $^4{\rm Li}$, we write Eq.~(\ref{corr-fun-def}) as 
\be
\label{yield-1}
\frac{dN_{p{^3{\rm He}}}}{d^3p_p  d^3p_{^3{\rm He}}} = \mathcal{R}({\bf p}_p , {\bf p}_{^3{\rm He}}) \, 
\frac{dN_p}{d^3p_p}  \frac{dN_{^3{\rm He}}}{d^3p_{^3{\rm He}}} ,
\ee
where the probability densities $\frac{dP_p}{d^3p_p}$,  $ \frac{dP_{^3{\rm He}}}{d^3p_{^3{\rm He}}}$ and $\frac{dP_{p{^3{\rm He}}}}{d^3p_p  d^3p_{^3{\rm He}}}$ are replaced by the particle number distributions $\frac{dN_p}{d^3p_p}$,  $\frac{dN_{^3{\rm He}}}{d^3p_{^3{\rm He}}}$ and $\frac{dN_{p{^3{\rm He}}}}{d^3p_p  d^3p_{^3{\rm He}}}$. 

We introduce the total momentum of the $p\!-\!^3{\rm He}$ pair, which is ${\bf P} \equiv {\bf p}_p + {\bf p}_{^3{\rm He}}$, and the relative momentum ${\bf q} \equiv \frac{1}{4}(3{\bf p}_p - {\bf p}_{^3{\rm He}})$. The correlation function strongly depends on ${\bf q}$ but the dependence of the product $\frac{dN_p}{d^3p_p}  \frac{dN_{^3{\rm He}}}{d^3p_{^3{\rm He}}}$ on ${\bf q}$ is rather weak in the momentum domain of the resonance. Therefore, the formula (\ref{yield-1}) can be written as
\be
\label{yield-2}
\frac{dN_{p{^3{\rm He}}}}{d^3q  d^3P} = \mathcal{R}({\bf q}) \, 
\frac{dN_p}{d^3p_p}  \frac{dN_{^3{\rm He}}}{d^3p_{^3{\rm He}}}\bigg|_{{\bf q}=0} ,
\ee
where the product  $\frac{dN_p}{d^3p_p}  \frac{dN_{^3{\rm He}}}{d^3p_{^3{\rm He}}}$ is taken at ${\bf q}=0$. 

\begin{figure}[t]
\begin{minipage}{84mm}
\centering
\includegraphics[scale=0.28]{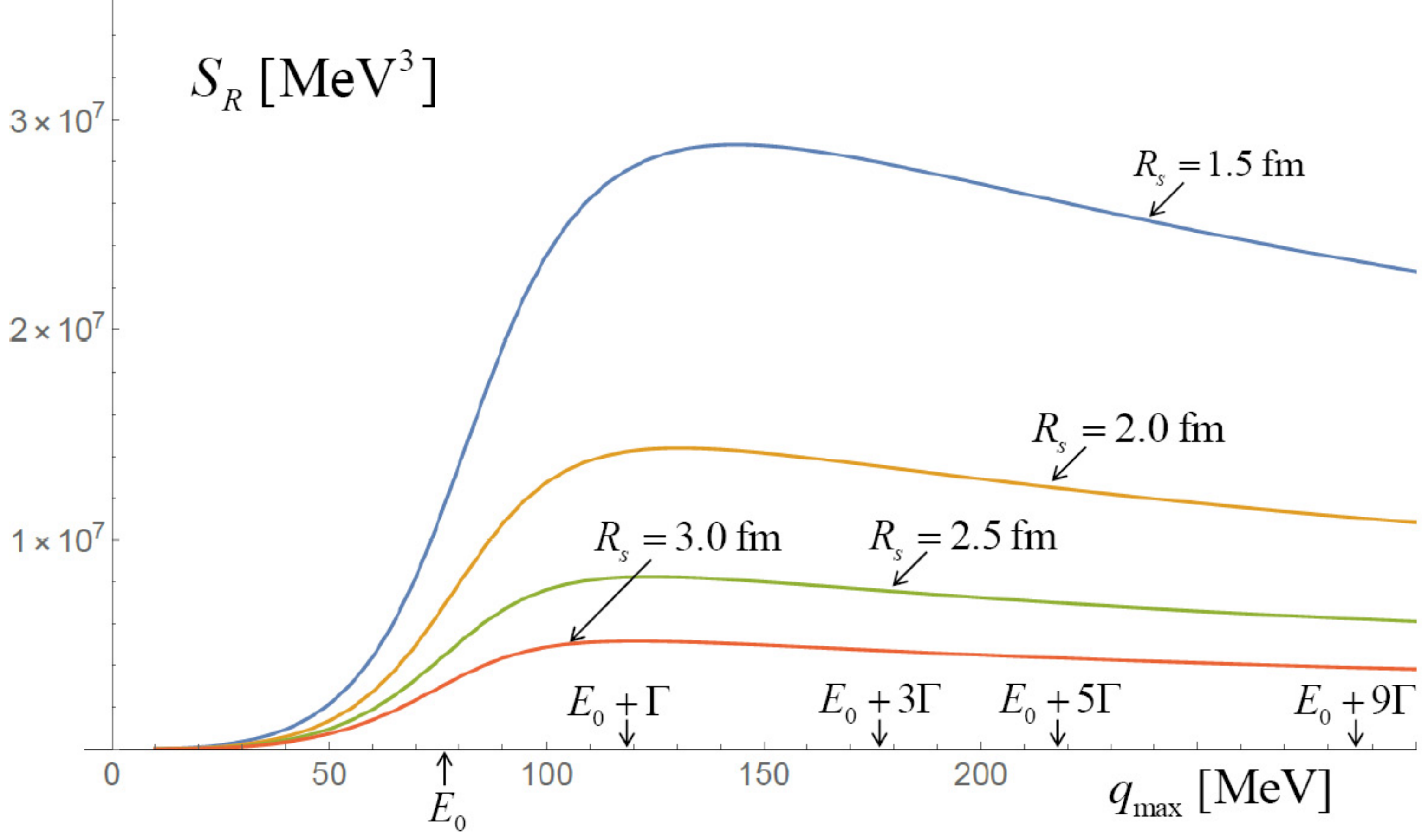}
\vspace{-6mm}
\caption{The quantity $S_R$ as a function of $q_{\rm max}$ for four values of $R_s$.}
\label{Fig-sR-qmax}
\end{minipage}
\hspace{6mm}
\begin{minipage}{84mm}
\centering
\vspace{-3mm}
\includegraphics[scale=0.22]{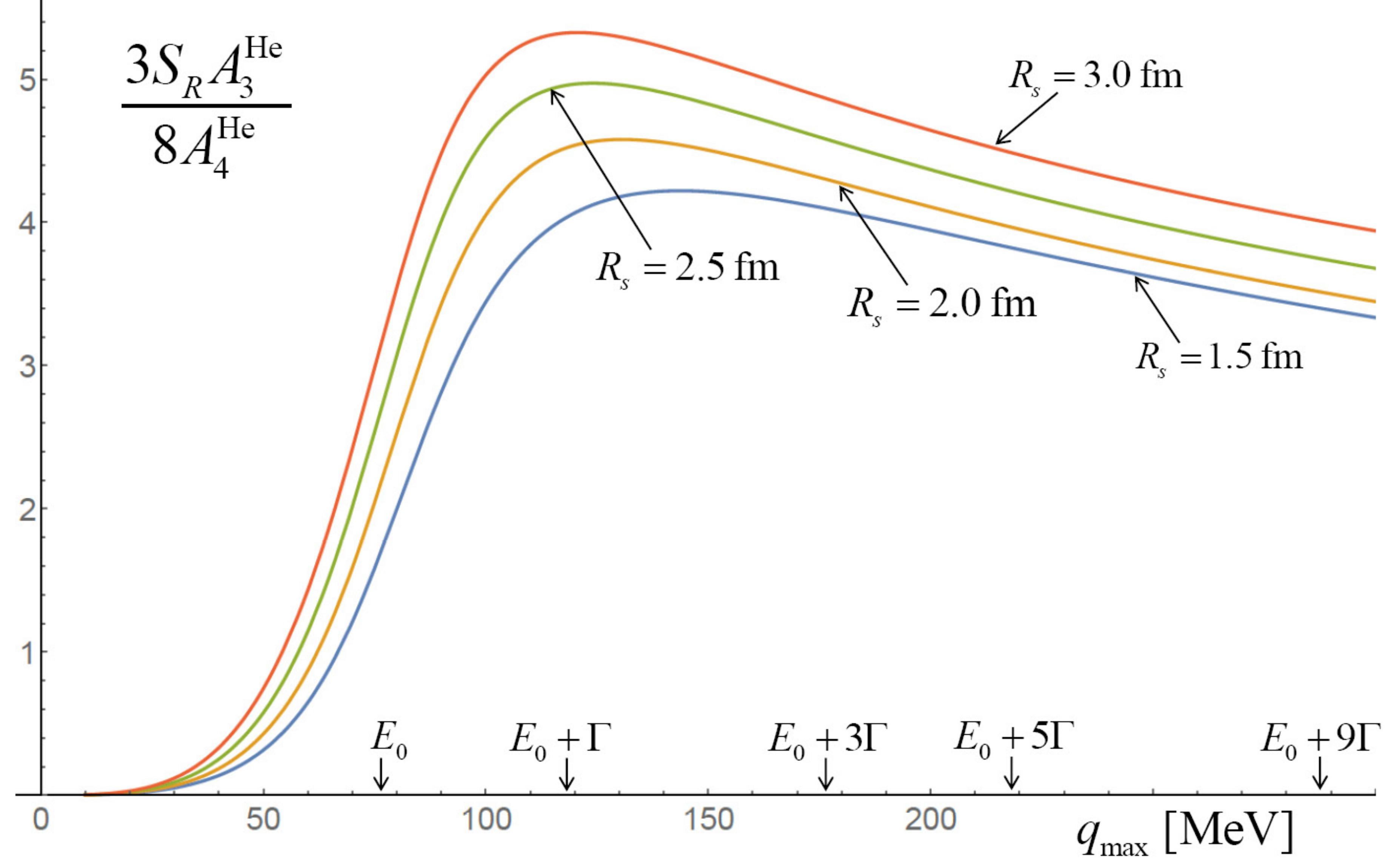}
\vspace{-2mm}
\caption{The ratio of $^4{\rm Li}$ to $^4{\rm He}$ yields as a function of $q_{\rm max}$ for four values of $R_s$.}
\label{Fig-ratio-qmax}
\end{minipage}
\end{figure}

To get the yield of $^4{\rm Li}$, one should sum up the number of correlated $p\!-\!^3{\rm He}$ pairs within the resonance peak. However, the peak is deformed by the Coulomb repulsion and $s-$wave scattering. So, we suggest to fit an experimentally obtained correlation function with the theoretical formula (\ref{fun-corr-gen-final}) where $\lambda_R$, which enters the amplitude (\ref{B-W-amplitude}) to control a strength of the resonance, is treated as a free parameter. Then, the contribution from the resonance can be disentangled. Denoting the correlation function shown in Fig.~\ref{Fig-resonance-1}, which differs from unity solely due to the resonance interaction, as $\mathcal{R}_R({\bf q})$, the yield of $^4{\rm Li}$ of the momentum ${\bf P}$ equals 
\be
\label{yield-final}
\frac{dN_{{^4{\rm Li}}}}{d^3P} = \frac{3}{4} \, S_R 
\frac{dN_p}{d^3p_p}  \frac{dN_{^3{\rm He}}}{d^3p_{^3{\rm He}}}\bigg|_{{\bf q}=0} ,
\ee
where the factor $3/4$ takes into account that  $^4{\rm Li}$ is produced only in the triplet channel and
\be
\label{sR-sR-qmax} 
S_R \equiv 4\pi \int_0^{q_{\rm max}} \Big(\mathcal{R}_R(q) - 1\Big) \,q^2 dq.
\ee
Since the source function is assumed to be isotropic and the correlation function depends on ${\bf q}$ only through $q$, the trivial angular integration has been performed in Eq.~(\ref{sR-sR-qmax}). The  upper limit of the integral (\ref{sR-sR-qmax}) should be chosen in a such way that the integral covers the resonance peak centered at $q_0 \approx 76~~{\rm MeV}/c$. As we already noted, our treatment of the resonance amplitude is not accurate beyond the vicinity of the resonance peak. However, the inaccuracy in the domain of small $q$, does not influence of the integral $S_R$ because the integrand is suppressed by the Jacobian $q^2$ in this domain. 

In Fig~\ref{Fig-sR-qmax} we show $S_R$  as a function of $q_{\rm max}$ for $\lambda_R = 1$ and four values of $R_s$. There are indicted the values of relative momenta of $^3{\rm He}$ and $p$ (in the center-of-mass frame) which correspond to the energy of the resonance peak $E_0$, to $E_0 + \Gamma$, to $E_0 + 3\Gamma$, etc. Fig~\ref{Fig-sR-qmax} shows that the integral (\ref{sR-sR-qmax}) changes rather slowly for $q_{\rm max}$ bigger than, say, 150 MeV but it is not clear whether the integral saturates when $q_{\rm max} \to \infty$. As observed in Ref.~\cite{Maj:2004tb} and further studied in \cite{Maj:2019hyy}, the analogous integrals of correlation functions usually diverge as $q_{\rm max} \to \infty$ because the correlation functions tend to unity as $q^{-3}$ or slower. However, it is not physically reasonable to extend the integral (\ref{sR-sR-qmax}) to a value of $q_{\rm max}$ higher than, say, $q_{\rm max} = 177$ MeV which corresponds to $E_0 + 3\Gamma$. The value of $S_R$ does not change very much when $q_{\rm max}$ is increased from 177 MeV to 286 MeV with the latter value corresponding to $E_0 + 9\Gamma$. 

To get the ratio of the yields of $^4{\rm Li}$ to $^4{\rm He}$ one has to express the yields of $^3{\rm He}$ and of protons through the yields of nucleons. Keeping in mind the coalescence formula (\ref{A-mom-dis}), Eq.~(\ref{yield-final}) is written as 
\be
\label{yield-final-N}
\frac{dN_{{^4{\rm Li}}}}{d^3P} = \frac{3}{8} \, S_R \, {\cal A}_3
\bigg(\frac{dN_N}{d^3p_N} \bigg)^4 ,
\ee
where the additional factor $1/2$ takes into account that half of nucleons are protons. Consequently, the ratio of $^4{\rm Li}$ to $^4{\rm He}$ yields equals
\be
\frac{{\rm Yield}(^4{\rm Li})}{{\rm Yield}(^4{\rm He})} = \frac{3 \, S_R \, {\cal A}_3}{8{\cal A}_4^{\rm He}} ,
\ee
and it is shown in Fig.~\ref{Fig-ratio-qmax} as a function of $q_{\rm max}$ for $\lambda_R = 1$ and four values of $R_s$. One sees that for $\lambda_R = 1$ and $q_{\rm max} \approx 150$ MeV the ratio varies between 4.0 and 5.5.

\section{Summary and Conclusions}
\label{sec-conclusions}

We propose to measure the yield of $^4{\rm Li}$ and compare it to that of $^4{\rm He}$ to falsify either the thermal or coalescence model which both properly describe yields of light nuclei produced in relativistic heavy-ion collisions at LHC. The nuclides $^4{\rm Li}$ and $^4{\rm He}$ have spins 2 and 0, respectively, while their masses are almost equal. Therefore, the ratio of their yields in the thermal model equals about 5 which reflects the different numbers of spin states of the two nuclides. 

In the coalescence model the yield of a nucleus depends on its internal structure. Since  $^4{\rm Li}$ is weakly bound and loose while $^4{\rm He}$ is well bound and compact the ratio of yields of $^4{\rm Li}$ to $^4{\rm He}$ is significantly smaller than 5 and it strongly increases when the source radius grows. Consequently, the ratio in the coalescence model depends, in contrast to that in the thermal model, on the collision centrality. 

The nuclide $^4{\rm Li}$ is unstable and it decays into $^3{\rm He}$ and $p$. Therefore, the yield of $^4{\rm Li}$ is accessible through a measurement of the $p\!-\!^3{\rm He}$ correlation function.  We have computed the function taking into account the resonance interaction responsible for the $^4{\rm Li}$ nuclide together with the $s-$wave scattering and Coulomb repulsion which significantly deform the resonance peak. Consequently, it is not evident how to infer the yield of the resonance from the correlation function. We propose to fit the experimentally obtained correlation function with the theoretical one where the resonance strength is a free parameter. Then, using Eq.~(\ref{yield-final}) the yield of $^4{\rm Li}$ at a given momentum can be obtained once the yields of  $^3{\rm He}$ and $p$  are known at the appropriate momenta. 

The  $p\!-\!^3{\rm He}$ correlation function carries the information encoded in a magnitude of the source radius whether $^3{\rm He}$ is emitted directly from the source or it is formed afterwards due final state interactions. If the source radius is accurately inferred from the  $p\!-\!p$ correlation function and the $p\!-\!^3{\rm He}$ correlation function is precisely measured, the information will be accessible. The  measurement is challenging but not impossible.

\section*{Acknowledgments}

We are grateful to Thomas Neff for a discussion on parity of nuclear states. This work was partially supported by the National Science Centre, Poland under grant 2018/29/B/ST2/00646. 


\end{document}